\documentclass{aa}  
\usepackage{graphicx}
\usepackage{txfonts}
\usepackage{natbib}
\bibpunct{(}{)}{;}{a}{}{,}
\newcommand{\doce}{\mbox{$^{12}$CO}}

\newcommand{\trece}{\mbox{$^{13}$CO}}

\newcommand{\jsc}{\mbox{$J$=6$-$5}}
\newcommand{\jtd}{\mbox{$J$=3$-$2}}
\newcommand{\jdu}{\mbox{$J$=2$-$1}}
\newcommand{\juc}{\mbox{$J$=1$-$0}}
\newcommand{\jdn}{\mbox{$J$=10$-$9}}
\newcommand{\jdq}{\mbox{$J$=16$-$15}}

\newcommand{\kms}{\mbox{km\,s$^{-1}$}}

\newcommand{\ms}{\mbox{$M_{\mbox{\sun}}$}}
\newcommand{\ls}{\mbox{$L_{\mbox{\sun}}$}}

\newcommand{\Tmb}{\mbox{$T_{\rm mb}$}}

\newcommand{\lsim}{\raisebox{-.4ex}{$\stackrel{\sf <}{\scriptstyle\sf \sim}$}}
\newcommand{\gsim}{\raisebox{-.4ex}{$\stackrel{\sf >}{\scriptstyle\sf \sim}$}}

\begin{document}
   \title{Extended rotating disks around post-AGB stars
}

\titlerunning{Extended rotating disks around post-AGB stars}

   \author{
          V. Bujarrabal\inst{1} \and J.\ Alcolea\inst{2} \and H. Van
          Winckel\inst{3} \and M.\ Santander-Garc\'{i}a\inst{2,4} \and
          A.\ Castro-Carrizo\inst{5} 
 }

   \offprints{V. Bujarrabal}

   \institute{             Observatorio Astron\'omico Nacional (OAN-IGN),
              Apartado 112, E-28803 Alcal\'a de Henares, Spain\\
              \email{v.bujarrabal@oan.es}
\and
             Observatorio Astron\'omico Nacional (OAN-IGN),
             C/ Alfonso XII, 3, E-28014 Madrid, Spain
\and
Instituut voor Sterrenkunde, KU Leuven, Celestijnenlaan 200D, 2001
Leuven, Belgium
\and
Centro de Astrobiolog\'{\i}a (CSIC-INTA), Ctra. M-108, km. 4,
E-28850 Torrej\'on de Ardoz, Madrid, Spain 
\and
 Institut de Radioastronomie Millim\'etrique, 300 rue de la Piscine,
 38406, Saint Martin d'H\`eres, France  
           }

   \date{July 2013, accepted}

  \abstract 
{There is a group of binary post-AGB stars that show a conspicuous NIR
  excess, usually assumed to arise from hot dust in very compact
  possibly rotating disks. These stars are surrounded by significantly
  fainter nebulae than the $standard$, well studied protoplanetary
  and planetary nebulae (PPNe, PNe), and, at least some of them, have
  significantly lower luminosities.
}
  {To identify and study extended rotating disks around these
    stars, shedding light on the role of disks in the
    formation and shaping of planetary nebulae.}
{We present high-sensitivity mm-wave observations of CO lines in 24
  objects of this type. The resulting CO lines are compared with
  profiles expected to arise from rotating disks, both from theoretical
  and observational grounds. We derive simple formulae that allow
  determining the mass of the CO-emitting gas and estimating its
  extent; the reliability and uncertainty of the methods are widely
  discussed.}
{CO emission is detected in most observed sources and the line
    profiles show that the emissions very probably come from disks in
    rotation. We derive typical values of the disk mass between
  10$^{-3}$ and 10$^{-2}$ \ms, about two orders of magnitude smaller
  than the (total) masses of $standard$ PPNe. The high-detection rate
  (upper limits being in fact not very significant) clearly confirm
  that the NIR excess of these stars arises from compact disks in
  rotation, very probably the inner parts of those found here.
  Low-velocity outflows are also found in about eight objects, with
  moderate expansion velocities of $\sim$ 10 \kms, to be compared with
  the velocities of about 100 \kms\ often found in $standard$
  PPNe. Except for two sources with complex profiles, the outflowing
  gas in our objects represents a minor nebular component. Our simple
  estimates of the disk typical sizes yields values $\sim$ 0.5 -- 1
  arcsec, i.e.\ between 5 10$^{15}$ and 3 10$^{16}$ cm.
Estimates of the linear momenta carried by the outflows, which can only
be performed in a few well studied objects, also yield moderate values,
compared with the linear momenta that can be released by the stellar
radiation pressure (contrary, again, to the case of the very massive
and fast bipolar outflows in $standard$ PPNe, that are strongly
overluminous). The mass and dynamics of nebulae around various classes
of post-AGB stars differ very significantly, and we can expect the
formation of PNe with very different properties. 
}
{}
\maketitle

\section{Introduction: low- and high-mass nebulae around post-AGB stars}

Planetary and protoplanetary nebulae (PNe, PPNe) very often show
axisymmetric shapes and fast axial movements, which are thought to be
due to shock interaction between the very collimated post-AGB jets and
the slow and isotropic AGB wind (see Balick \& Frank 2002 and
references therein). Many PPNe already contain most of the initial
stellar mass (in some way, the star has become a nebula), and a
significant fraction of this mass is accelerated by these shocks, up to
typically 50 -- 200 \kms\ (see e.g.\ Bujarrabal et al.\ 2001). We note
that the best studied objects tend to be relatively massive and
luminous, because of the usual observational bias.  Most of the gas in
most of these {\em standard} PPNe is in molecular form. However, in
more evolved objects the effects of photodissociation due to the hot
central star are already noticeable and the gas is often composed of
neutral or ionized atoms.  In the final stage of planetary nebulae,
many nebulae are found to contain in total $\sim$ 0.2 -- 1
\ms\ (Huggins et al.\ 1996, Kimura et al.\ 2012, Castro-Carrizo et
al.\ 2001, Fong et al.\ 2001), in the form of atomic or molecular gas.

In well studied (massive) nebulae, the formation of these axial
structures involves enormous amounts of energy and linear momentum,
well above what the stellar radiation can provide. On the other hand,
the main phase of the PN shaping, when the shocks are actually
accelerating the nebular material, is extremely short, lasting just a
few hundred years (Bujarrabal et al.\ 2001; in total, the whole
protoplanetary phase is thought to last $\sim$ 1000 yr, and the typical
lifetime of the PN phase, when the nebula just follows a ballistic
expansion, is $\sim$ 10000 yr).

The presence of gas disks orbiting post-AGB stars is often postulated
to explain the very energetic PPN dynamics and the collimation of the
post-AGB jets (see e.g.\ Soker 2002; Frank \& Blackman 2004). Mass
re-accretion from rotating disks (in the presence of a magnetic field)
can provide the required energy and momentum, in a process similar to
that at work in forming stars. In the case of evolved stars, the
presence of a stellar or substellar companion is required to explain
the extra angular momentum that is necessary to form a rotating disk
from previously ejected material, since it is expelled with negligible
angular momentum.  The presence of rotating disks around post-AGB stars
and their main properties is, therefore, a basic question in order to
understand the post-AGB ejections, and therefore the spectacular and
very fast evolution of the shape and dynamics of young PNe.

We recall that torus-like or oblate structures of molecular gas are
commonly detected in the central equatorial regions of protoplanetary
and young planetary nebulae, although they are not observed to rotate,
but to be systematically in expansion (like the rest of the
nebula). Such expanding structures are usually thought to be mere
remnants of the former AGB winds.  See the cases of M\,1--92 (Alcolea
et al.\ 2007), M\,2--9 (Castro-Carrizo et al.\ 2012), M\,2--56
(Castro-Carrizo et al.\ 2002), etc.

Not all post-AGB stars are surrounded by such thick shells.
There is a population of faint (evolved) PNe in which only
significantly smaller amounts of mass ($<$ 0.1 \ms) have been detected
(Acker et al.\ 2012, Huggins et al.\ 1996, Pottasch 1984), perhaps
because most of the mass is placed in extended low-brightness haloes or
simply due to the lack of nebular material. These objects are less
studied than more massive PNe.

There is also a group of early post-AGB objects in which the stellar
component dominates and are surrounded by low-mass nebulae.  These
objects are old, yellow, low-gravity, relatively luminous stars (but in
general less luminous than the massive {\em standard} PPNe presented in
previous paragraphs). Most of them also show relatively low stellar
temperatures, usually $<$ 10000 K.  According to these properties, they
are thought to have evolved from a previous red giant phase but to be
still far from the blue or white dwarf stage (see e.g.\ van Winckel
2003). Contrary to the case of massive protoplanetary nebulae, most
of these stars are bright in the optical since the circumstellar
extinction is not high, and so they are often identified from their
stellar properties.

All these optically-bright post-AGB objects show a clear FIR excess,
since some nebular material is in any case present.  A significant
fraction of them also show a remarkable NIR excess, which has been
attributed to the emission of hot dust, at about 1000 K. The high
temperature would show that dust grains are kept close to the star, in
spite of the expected low mass-loss rate after the end of the AGB and
the slow evolution of these sources. A systematic search for bright
post-AGB stars (see van Winckel et al.\ 2009, van Aarle et al.\ 2012,
and references therein) resulted in catalogs in which about half of the
candidates show a NIR excess of this kind; among these are the
RV\,Tauri pulsators. Often these objects show a peculiar photospheric
composition which is depleted in refractory elements. This is likely
due to reaccretion of only the gas from the dusty circumstellar
environment (Waters et al.\ 1992). Another observational characteristic
of these objects is that the infrared dust spectra are best interpreted
assuming quite large grains with a very high crystallinity fraction
(Gielen et al., 2011).  All these observational characteristics suggest
that these stars are surrounded by very compact and stable nebular
components, likely rotating discs, see e.g.\ Van Winckel (2003).  The
compact nature of the NIR emitting component is confirmed by
interferometric experiments (Deroo et al.\ 2007), which also suggested
an elongated, disk-like shape. Practically all the stars of this kind
(bright post-AGB stars with NIR excess) that have been studied in
detail are close binaries (e.g.\ Van Winckel et al.\ 1999, de Ruyter et
al.\ 2005), which would explain the origin of the angular momentum
required to form the disks. The exceptions could still be objects in
which the study of the radial velocity variations is hampered by the
stellar pulsation.

However, the rotation dynamics of these small disks is not yet well
established, because no spectroscopic data has been able to probe their
velocity field.  We cannot discard that the presence of hot dust is due
to a very recent mass ejection or expansion at a very low
velocity. Indeed, very slow outflowing gas was found in the very inner
regions of some {\em standard} PPNe, probably ejected during the
very last AGB phases (S\'anchez-Contreras et al.\ 2004, Castro-Carrizo
et al.\ 2012).

Among these objects with significant NIR excess, we can mention the
variable stars of RV Tau type, the Red Rectangle, 89 Her, HR\,4049, etc
(see e.g.\ Trams et al.\ 1991, Alcolea \& Bujarrabal 1991, Waters et
al. 1992, Van Winckel 2003).  In the Red Rectangle and 89 Her, probably
the best studied of these sources, the total nebular mass is
\lsim\ 0.01 \ms\ (e.g.\ Bujarrabal et al.\ 2001, 2005, 2007). The other
objects, in which the studies of the nebular component are less
detailed, show a nebular mass always smaller than 0.1 \ms, see Alcolea
\& Bujarrabal (1991), Van Winckel (2003), etc. We cannot exclude the
presence of outer haloes or very inner components with significant
amounts of mass (Men'shchikov et al.\ 2002), but these are difficult to
detect.

Only in one object, the Red Rectangle, the presence of an extended disk
in rotation has been demonstrated, by means of interferometric maps of
mm-wave CO lines (Bujarrabal et al.\ 2005).
On the other hand, the detailed analysis of Herschel observations of
FIR high-$J$ lines by Bujarrabal et al.\ (2013) suggested, in order to
explain the relatively intense line-wing emission in this object, the
presence of diffuse gas expanding at about 10 \kms, probably in the
form of axial blobs.

89 Her was also mapped in low-$J$ CO emission (Bujarrabal et
al.\ 2007). The nebula is dominated by an hour-glass like structure
about 8$''$ wide and expanding at about 5 \kms. There is also a central
intense component, practically unresolved in the observations
(\lsim\ 1$''$), which  was suggested to be in rotation in view of its
small velocities and extent.

Other previously published CO observations of these post-AGB stars
yielded less detailed results (see Alcolea \& Bujarrabal 1991,
Bujarrabal et al.\ 1988, 1990, and Maas et al.\ 2003), and only two
more objects were detected.  R Sct (a peculiar RV Tau variable) shows a
narrow but complex single-dish profile.  IRAS\,08544-4431 showed
relatively narrow profiles with a central peak similar to those found
in 89 Her, which could indicate emission from a rotating disk.  AC Her
(another RV Tau star) may show a narrow \doce\ \jdu\ line, but it is
weak and was just tentatively detected. Other similar objects were
observed but not detected in those observations.

Is the extended rotating disk of the Red Rectangle a very peculiar
characteristic of this nebula?
Or, on the contrary, are rotating disks present in many other post-AGB
objects? Are rotating equatorial disks and expanding axial blobs
simultaneously present? (at least in low-mass post-AGB nebulae). And
finally, even if there are rotating disks in these low-mass nebulae,
which require moderate energies to be formed, can we invoke disks to
explain the dynamics of massive nebulae, in which we do not find any
observational sign of the presence of rotating structures?  This paper
is devoted to throw some light on these questions.

\begin{table*}[bthp]
\begin{center}                                          
\caption{Post-AGB sources observed with the 30m IRAM telescope,
  observed coordinates, and some other relevant properties: IRAS
  fluxes, total luminosity and adopted distance (Sect.\ 2.1), 
  stellar spectral type, and stellar
  velocity (derived from optical
  lines, except for sources previously detected in CO emission).}

\scriptsize

\vspace{-.3cm}
\begin{tabular}{|l|cccccccc|l|}
\hline\hline
& & &  &  &  &  & & & \\ 
Source & \multicolumn{2}{c}{observed coordinates} & $F_{25\mu}$ &
$F_{60\mu}$ & luminosity  &  distance  & sp.\ type  & $V_*$(LSR) & comments \\
& \multicolumn{2}{c}{J2000} & Jy & Jy & \ls & pc & & \kms & \\ 
\hline
 & & &  &  &  &  & & & \\ 
RV Tau          & 04:47:06.73  &  +26:10:45.6    & 18.1 &  6.5 & 3500 
& 2200$^1$   & G2-M2 & +17 & RV Tau variable \\
DY Ori          & 06:06:14.91  &  +13:54:19.1    & 14.9  &  4.2 & 1600 &
2000$^1$  & G0 I & --16 & RV Tau variable \\
Red Rectangle   & 06:19:58.22  & --10:38:14.7    & 456.1 & 173.1 & 6000
& 710$^2$   & B9 & 0 & interferometric CO maps \\
U Mon            & 07:30:47.47  & --09:46:36.8    & 88.4  & 26.6 & 4000 
& 800$^1$ & F8-K0 I & +9 & RV Tau variable \\
AI CMi          & 07:35:41.15  &  +00:14:58.0    & 68.1  &  18.5 &
3000$^3$ & 1500$^3$ & G5 I & +33 & \\
HR 4049         & 10:18:07.59  & --28:59:31.2    &  9.6  &   1.8 & 2900 &
650$^4$  & A6 I & --45 & well studied inner disk \\
89 Her          & 17:55:25.19  &  +26:03:00.0    & 54.5  &  13.4 & 9200
& 1000$^5$ & F2 I & --8 & interferometric CO maps\\
IRAS 18123+0511 & 18:14:49.39  &  +05:12:55.7    & 11.0  &   4.2 &
3000$^3$ & 3500$^3$ & F2 & +99 &  \\  
AC Her          & 18:30:16.24  &  +21:52:00.6    & 65.3  &  21.4 & 3500
& 1100$^1$ & F2-K4 I & --10 & previous tent.\ det., RV Tau variable \\
R Sct           & 18:47:28.95  & --05:42:18.5    &  9.3  &  8.2 & 14000 
& 1000$^5$ & G0-K0 I & +55 &  previous det., pec.\ RV Tau variable \\    
IRAS 19125+0343 & 19:15:01.18  &  +03:48:42.7    & 26.5  &  7.8 & 3000$^3$
& 1500$^3$ & F2 & +84 & \\   
IRAS 19157-0247 & 19:18:22.71  & --02:42:10.9    & 7.2   &  2.5 & 3000$^3$
& 2900$^3$ & F3 & +49 & \\  
IRAS 20056+1834 & 20:07:54.62  &  +18:42:54.5    & 18.0  &  5.4 & 3000$^3$
& 3000$^3$ & G0 & $\sim$0 & edge-on disk \\   
R Sge           & 20:14:03.75  &  +16:43:35.1    & 7.5   &  2.1 & 2500 
& 2500$^1$ & G0-K0 I & +28 & RV Tau variable \\  
\hline\hline
\end{tabular}
\end{center}
Distance estimate: $^1$: P-L relation for RV Tau variables. $^2$:
Men'shchikov et al.\ (2002).  $^3$: Assumed luminosity.  $^4$: Corrected
parallax (Acke et al.\ 2013).  $^5$: Hipparcos parallax.

\end{table*}

\begin{table*}
\begin{center}                                          
\caption{Southern post-AGB sources observed with APEX telescope,
  observed coordinates, and some other relevant properties (see
  Table 1).}

\scriptsize

\vspace{-.3cm}
\begin{tabular}{|l|cccccccc|l|}
\hline\hline
& & &  &  &  &  & & & \\ 
Source & \multicolumn{2}{c}{observed coordinates} & $F_{25\mu}$ &
$F_{60\mu}$ & luminosity  &  distance  & sp.\ type  & $V_*$(LSR) & comments \\
 & \multicolumn{2}{c}{J2000}  & Jy & Jy & \ls & pc & & \kms & \\ 
\hline
 & & &  &  &  &  & &  & \\ 
AR Pup & 08:03:01.65  &  --36:35:47.9 & 94.3 & 26.1 & 3500 & 2800$^1$ 
& F0I & +29 & RV Tau variable \\
IRAS 08544-4431 & 08:56:14.18 & --44:43:10.7 & 158.8 & 56.3 & 3000$^2$ &
550$^2$  & F3 & +45 & previous det., pec.\ RV Tau variable \\ 
IW Car & 09:26:53.30 & --63:37:48.9 & 96.2 & 34.5 & 3000 & 1000$^1$  
& A4I & --25 & RV Tau variable \\ 
IRAS 10174-5704 & 10:19:16.89 & --57:19:26.0 & 60.3 & 16.1 & 3000$^2$ &
850$^2$  & G8I & +3 & \\
IRAS 10456-5712 & 10:47:38.40 & --57:28:02.7 & 115.5 & 30.9 & 3000$^2$ &
600$^2$  & M1 & --9 &
\\
HD 95767 & 11:02:04.31 & --62:09:42.8 & 15.7 & 10.9 & 3000$^2$ & 1400$^2$  
& F3II & --32 & \\
RU Cen & 12:09:23.81 & --45:25:34.8 & 11.0 & 5.7 & 2500 & 2200$^1$  
& G2 & --35 & RV Tau variable \\
HD 108015 & 12:24:53.50 & --47:09:07.5 & 33.23 & 8.0 & 3000$^2$ & 1800$^2$ 
& F4I & --2 & \\
IRAS 15469-5311 & 15:50:43.80 & --53:20:43.3 & 42.1 & 15.5 & 3000$^2$ &
1200$^2$ & F3 & --14 &  \\
IRAS 15556-5444 & 15:59:32.57 & --54:53:20.4 & 17.8 & 7.1 & 3000$^2$ &
1700$^2$ & F8 & ? &\\
\hline\hline
\end{tabular}
\end{center}
Distance estimate: $^1$: P-L relation for RV tau stars. $^2$: assumed
luminosity. 
\end{table*}

\section{Source sample}

We have selected 24 sources to be observed in CO mm-wave lines
following three criteria: 

\noindent
{\bf 1.} They have been identified as post-AGB stars, in view of their
evolved nature (not associated to interstellar clouds and the galactic
plane), low gravity, relatively high luminosity, intermediate spectral
type, and FIR excess (indicative of material ejected by the star). We
selected sources from the catalog by de Ruyter et al.\ (2006); we also
included low-mass post-AGB stars observed by Alcolea \& Bujarrabal
(1991).

\noindent
{\bf 2.} They all show NIR excess indicative of the presence of hot
dust, possibly placed close to the star (Sect.\ 1). The NIR excess is
very small in the case of R Sct, and particularly strong in
IRAS\,20056+1834. In disk sources the viewing angle of the disk is an
important parameter which will determine the SED. For IRAS\,20056+1834,
there is observational evidence that we look edge-on (Kameswara Rao et
al.\ 2002), and the stellar light is severely obscured.

\noindent
{\bf 3.} We selected intense sources in the FIR, to increase the CO
detection rate. The intense FIR would show that there is more extended
material, in which we could find molecular gas, and/or the distance is
not very large. A CO -- 60\,$\mu$m flux correlation is found in {\em
  standard} (massive) PPNe (e.g.\ Bujarrabal et al.\ 1992) and there is
also a hint of CO/FIR relation in the few previous observations of
similar objects (Alcolea \& Bujarrabal 1991).

\noindent
{\bf 4.} A first group of 14 sources was selected imposing that the
declinations cannot be very negative (over -25$^{\circ}$), in
order to be observed with the IRAM 30m telescope. A second group of 10
southern sources was independently selected, in order to be observed
with the APEX telescope.

The sources observed with the two telescopes, the IRAM 30m telescope
and the APEX telescope, are treated quite differently in this paper,
because the data obtained with APEX are more difficult to
interpret. The APEX data do not include \trece\ \juc\ lines, which are
the basic data in our analysis and deduction of the main disk
properties. Moreover, a significant fraction of the APEX observations
suffer strong contamination due to galactic ISM emission, yielding poor
limits to the disk emission. For these reasons, APEX data are more
useful to complement the list of post-AGB objects showing disk emission
than to estimate the detection rate and disk parameters.

The objects observed with the 30m and APEX telescopes are respectively
presented in Tables 1 and 2, where we give the observed coordinates,
the IRAS 25 and 60 $\mu$m fluxes, luminosities, distances, stellar
spectral types and velocities, and some comments. We have tried to give
the velocities of the stellar system center of gravity, by averaging
values from, if possible, many different observations; the
  uncertanity of the velocity determinations vary from object to
  object, with a typical value of $\sim$  5 \kms.

\subsection{Estimates of the distance and luminosity of the studied
  objects} 

Distances of the Galactic disk sources are poorly constrained as the
sources are generally too far for direct distance detection via the
parallax measurement. One of the exceptions is HR\,4049 (Acke et
al.\ 2013) for which the parallax estimate had to be corrected for the
orbital motion of the primary around the center of mass. We adopted
different strategies to estimate the distances of the other sources:
for the RV\,Tauri stars we used the P-L relation of Alcock et al (1998)
calibrated to the LMC. We used the detailed study of the LMC RV\,Tauri
pulsators of Gielen et al. (2009) to obtain better estimates of the
intrinsic luminosities. We then modelled the SED similar to what has
been performed in e.g. Gielen et al. (2011), using appropriate Kurucz
model atmospheres and broad-band photometric observations obtained
using Simbad. We dereddened the raw photometric data to obtain the
total line-of-sight reddening by $\chi^2$ minimization between the
dereddened photospheric fluxes and the atmosphere model. The
integration of the dereddened photospheric emission and the IR
luminosity was then used to obtain the distances.

For non-pulsating objects for which there is no indication on the
distance, we assumed a default luminosity of 3000 L$_{\odot}$ and
modelled the SED and determined the distance accordingly.
IRAS\,20056+1834 (Kameswara Rao et al., 2002, Gielen et al., 2011) is
mainly an IR source and the optical dereddened luminosity is dwarfed by
the total infrared excess. The likely reason is that we look at the
system edge-on and that the optical photons are dominated by scattered
light. We therefore used the full SED integral as a proxy for the total
luminosity and used this to constrain the distance. Another example in
which we look at the system nearly edge-on is the Red Rectangle, for
which the disk is resolved in the optical (Cohen et al., 2004); for the
Red Rectangle we took the distance estimated by Men'shchikov et
al.\ (2002) from their detailed modeling of the nebula emission. R\,Sct
is a peculiar RV\,Tauri pulsator whose true nature is still a matter of
debate (e.g. Matsuura et al., 2002). For this object we took the
parallax measurement which yields a distance of 1 kpc and obtained the
luminosity by integrating the SED. 

\section{Observations}

\subsection{IRAM 30m data}

Data for the \doce\ and \trece\ \jdu\ and \juc\ lines have been taken
with the IRAM 30m telescope.  The observations were performed in three
observing sessions, in May 2012, January 2013, and May 2013. We used
the new EMIR (Eight MIxer) receivers, able to observe simultaneously in
the 3\,mm and 1\,mm bands in dual polarization mode.  During the first
session we observed the four CO lines simultaneously, but only for the
\doce\ \jdu\ line the two polarizations could be recorded.  In the
second and third sessions, we also observed the three weakest lines,
\doce\ \juc\ and \trece\ \jdu\ and \juc\ in dual polarization. For
each line, observations in the two polarizations were averaged, after
checking that the line intensities were compatible.
The data were recorded using the FTS, with a resolution of 200 kHz,
equivalent to about 0.25 and 0.5 \kms\ in the 1mm and 3mm bands,
respectively. In some cases, namely for very weak and/or wide lines,
the spectral resolution was degraded by a factor 2 or 4, by averaging
adjacent channels, to better determine the line parameters.

The spatial resolution of the observations at 3\,mm is 22--23$''$, and
12--13$''$ at 1\,mm wavelength. Frequent pointing measurements were
performed to monitor and correct pointing errors; errors not larger
than 3$''$ were typically found, which have practically no effects on
the calibration. The observations were done by wobbling the
subreflector by 2$'$ at a 0.5\,Hz rate. This method is known to provide
very stable and flat spectral baselines. The atmospheric conditions
during the observations were good. The average zenith opacity during
the observing runs was \lsim\ 0.3, slightly better at 110 GHz and
slightly worse at 115 GHz.

The data presented here have been calibrated in units of
(Rayleigh-Jeans-equivalent) main-beam temperature, corrected for the
atmospheric attenuation, $T_{\rm mb}$, using the standard chopper wheel
method. Calibration scans (observation of the hot and cold loads, and
of the blank sky) have been performed typically every 15--20
minutes. The temperature scale is set by observing hot and cold loads
at ambient and liquid nitrogen temperatures. Correction for the antenna
coupling to the sky and other losses have been done using the latest
values for these parameters measured at the telescope. The sky
attenuation is computed from the values of a weather station, the
measurement of the sky emissivity, and a numerical model for the
atmosphere at Pico de Veleta.

The observed $T_{\rm mb}$ values have been re-scaled after the
observation of the well known emitters IRC\,+10216 (CW\,Leo) and
NGC\,7027, whose CO emission is strong and well characterized. For
IRC\,+10216 we adopted $T_{\rm mb}$ values at line center of 50 K, 36
K, 23 K, and 1 K, for respectively the \doce\ and \trece\ \jdu\ and the
\doce\ and \trece\ \juc\ transitions. For the profile peaks of the same
lines in NGC\,7027, we adopted the values 25 K, 1.05 K, 11.9 K, and
0.27 K. We also took into account previous observations of CRL\,618 and
89 Her (Bujarrabal et al.\ 2001); note that more recent observations
tend to indicate that the 1mm data in that paper are probably
overestimated by about a 20\%.  After these corrections, and taking
into account variations we have found between observations in these
runs as well as our experience with previous data, we believe that the
absolute incertitude in fluxes is approximately $\pm$20\%. 

Baselines of degree 1 to 3 were subtracted from our spectra. The
resulting line profiles are shown in figures 1 to 11 for the sources
in which emission was detected. The spectra for undetected objects are
presented in Appendix A. A summary of the observational parameters is
given in Table 3: velocity range occupied by the emission (or the whole
analyzed band for undetected sources), spectral resolution used to
determine the tabulated parameters, peak intensity, $rms$ noise, line
area and uncertainty, and equivalent width, as well as some
comments. The equivalent width is defined as the profile area divided
by the peak. The noise in the line area is estimated as the $rms$
noise multiplied by the adopted equivalent width and divided by
$\sqrt{n_{\rm ch}}$, where $n_{\rm ch}$ is the number of velocity
channels within the equivalent width (i.e.\ the equivalent width
divided by the considered spectral resolution); we note that in
some cases the uncertainty in the baseline retrieval can yield
noticeable (but difficult to quantify) errors in the line area.

The velocity ranges given in Table 3 for detected sources are those
occupied by the full line profile, used in particular to calculate the
line area. For nondetections, the velocity range is the total range
considered to estimate the upper limits, and are centered on the $LSR$
stellar velocities deduced from optical data (Tables 1, 2).  The $LSR$
peak velocity of the detected lines is found to be coincident with that
of previous detections or, in the case of new detections, with the
stellar velocity deduced from optical data (within $\pm$ 5 \kms). For
IRAS\,20056+1834, which shows a quite asymmetrical profile, the stellar
velocity deduced from the optical ($\sim$ 0 \kms) is uncertain, but
close to the velocity of the line peak, while the line-wings appear
blue-shifted.

   \begin{figure}
   \centering \rotatebox{0}{\resizebox{8.5cm}{!}{ 
\includegraphics{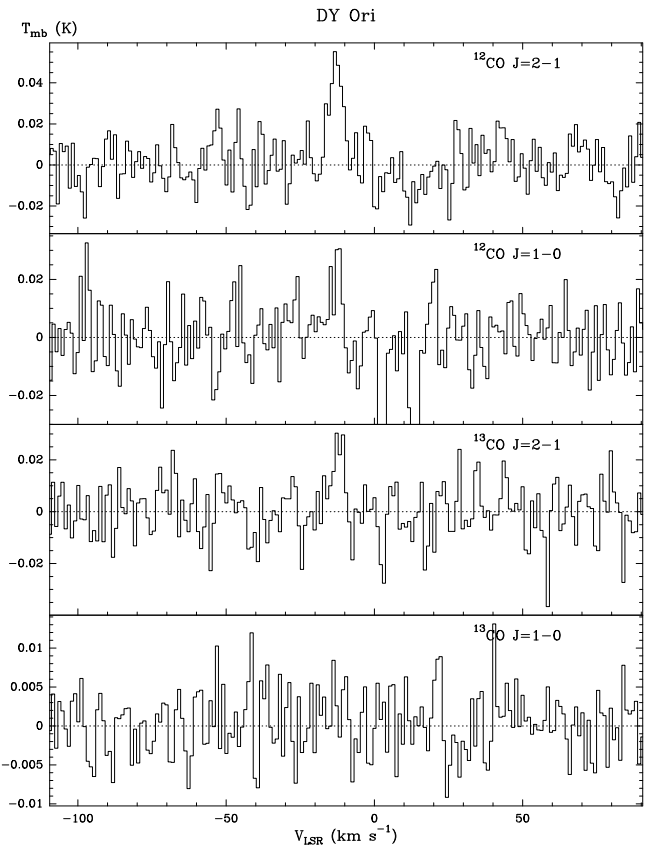}
}}
   \caption{Observed profiles in DY Ori. }
              \label{}%
    \end{figure}

   \begin{figure}
   \centering \rotatebox{0}{\resizebox{8.5cm}{!}{ 
\includegraphics{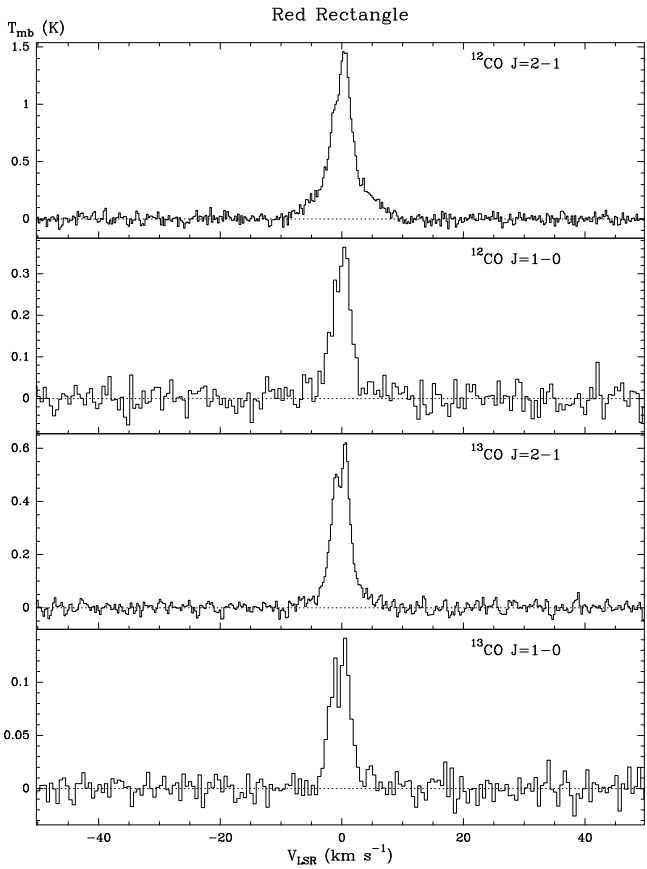}
}}
   \caption{Observed profiles in the Red Rectangle. }
              \label{}%
    \end{figure}

   \begin{figure}
   \centering \rotatebox{0}{\resizebox{8.5cm}{!}{ 
\includegraphics{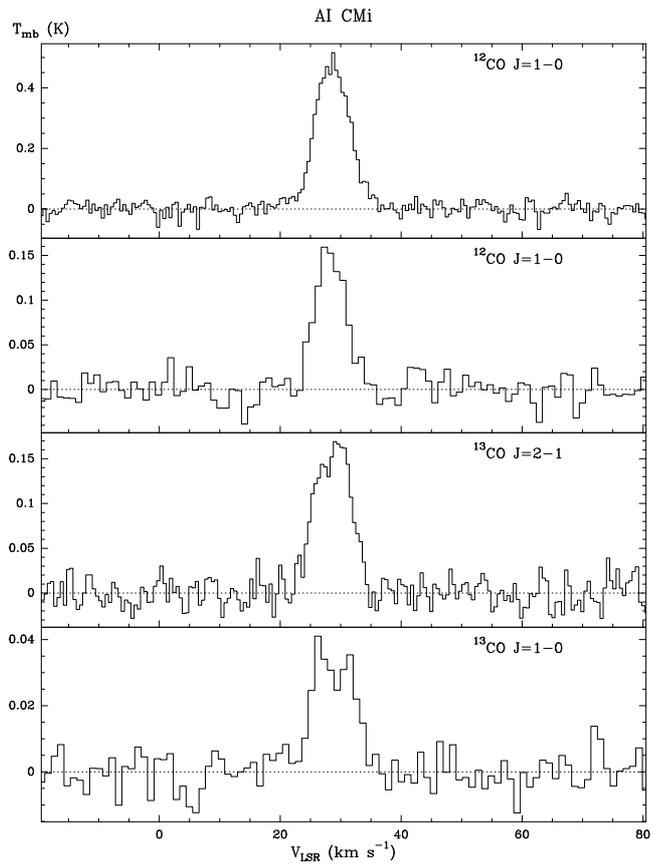}
}}
   \caption{Observed profiles in AI CMi. }
              \label{}%
    \end{figure}

   \begin{figure}
   \centering \rotatebox{0}{\resizebox{8.5cm}{!}{ 
\includegraphics{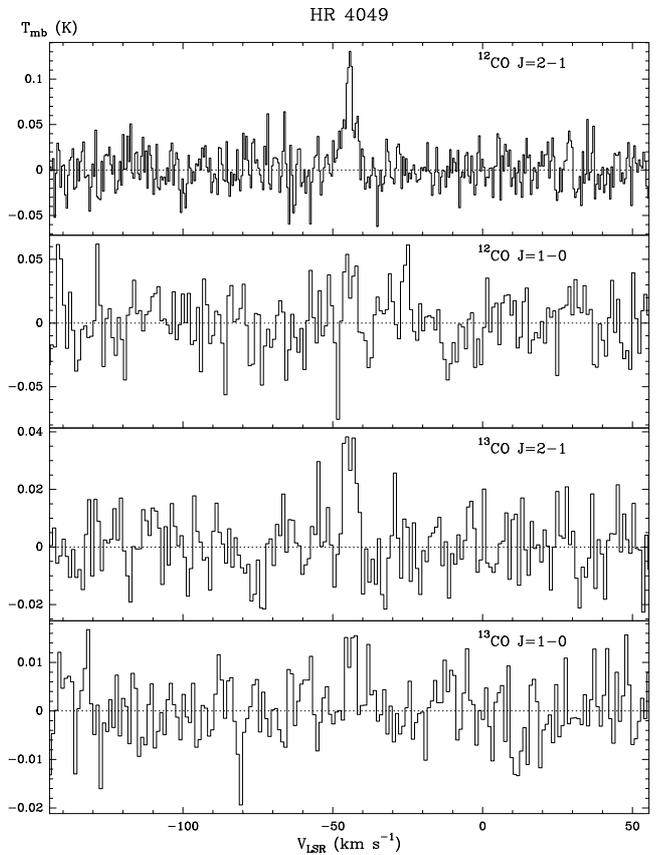}
}}
   \caption{Observed profiles in HR 4049. }
              \label{}%
    \end{figure}

   \begin{figure}
   \centering \rotatebox{0}{\resizebox{8.5cm}{!}{ 
\includegraphics{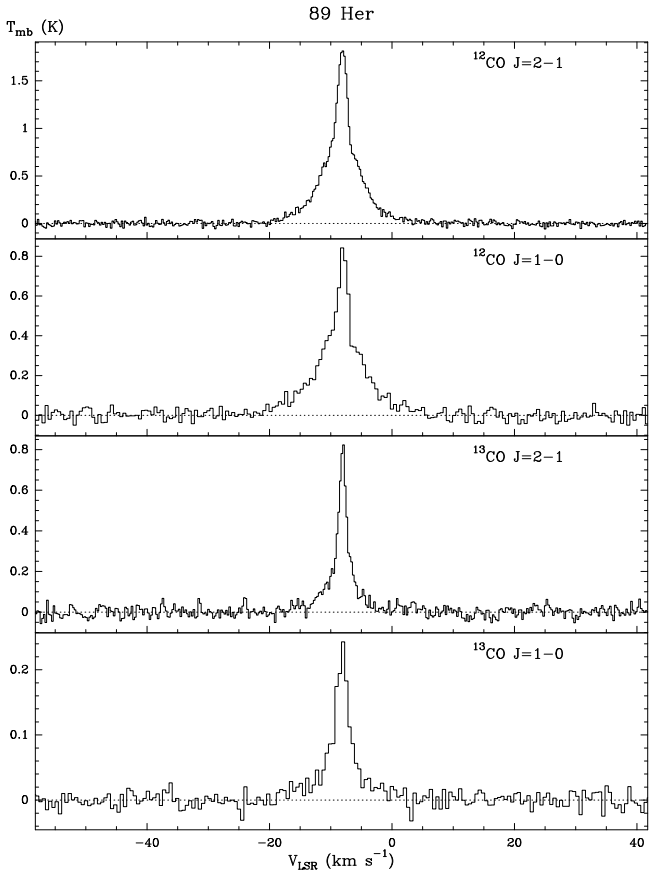}
}}
   \caption{Observed profiles in 89 Her. }
              \label{}%
    \end{figure}

   \begin{figure}
   \centering \rotatebox{0}{\resizebox{8.5cm}{!}{ 
\includegraphics{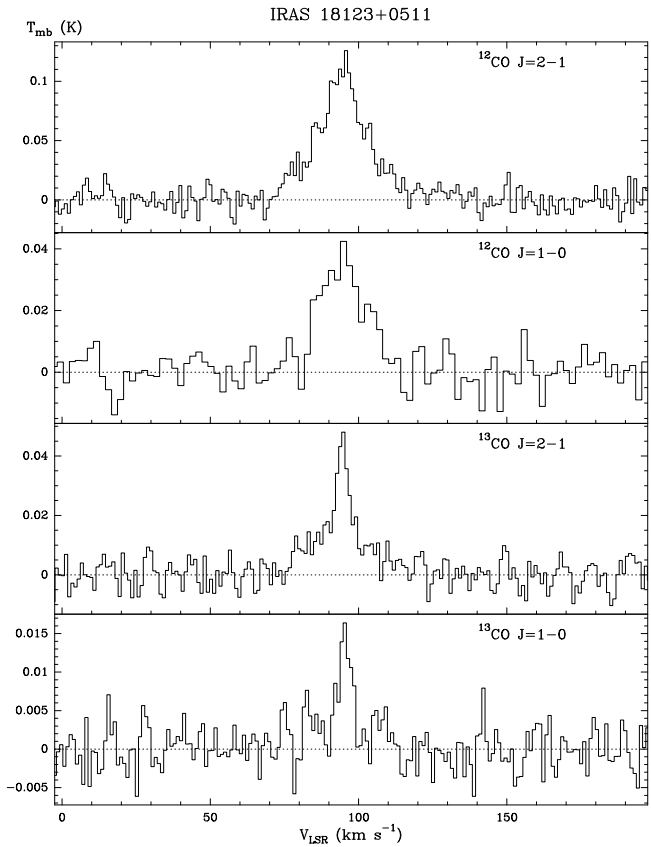}
}}
   \caption{Observed profiles in IRAS 18123+0511. }
              \label{}%
    \end{figure}

   \begin{figure}
   \centering \rotatebox{0}{\resizebox{8.5cm}{!}{ 
\includegraphics{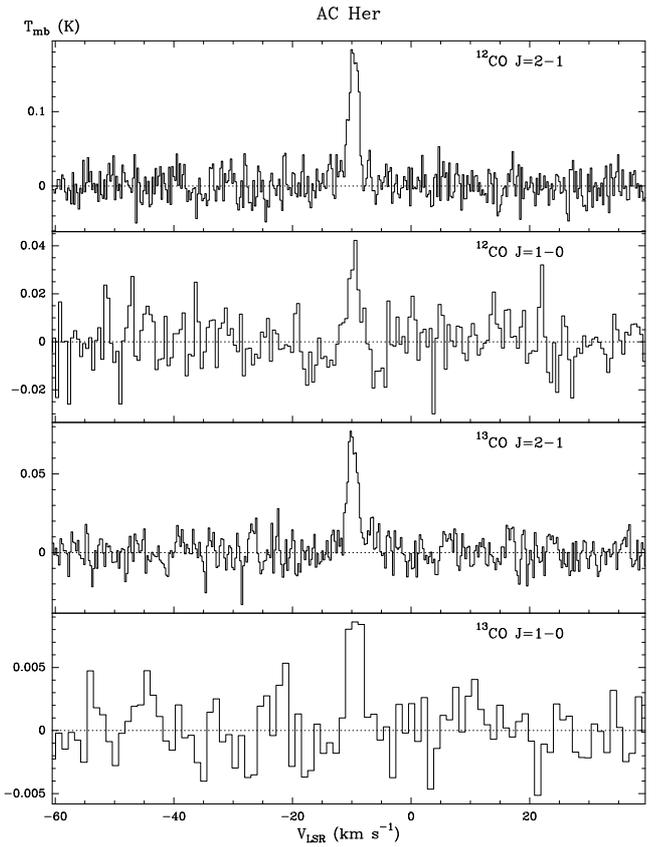}
}}
   \caption{Observed profiles in AC Her. }
              \label{}%
    \end{figure}

   \begin{figure}
   \centering \rotatebox{0}{\resizebox{8.5cm}{!}{ 
\includegraphics{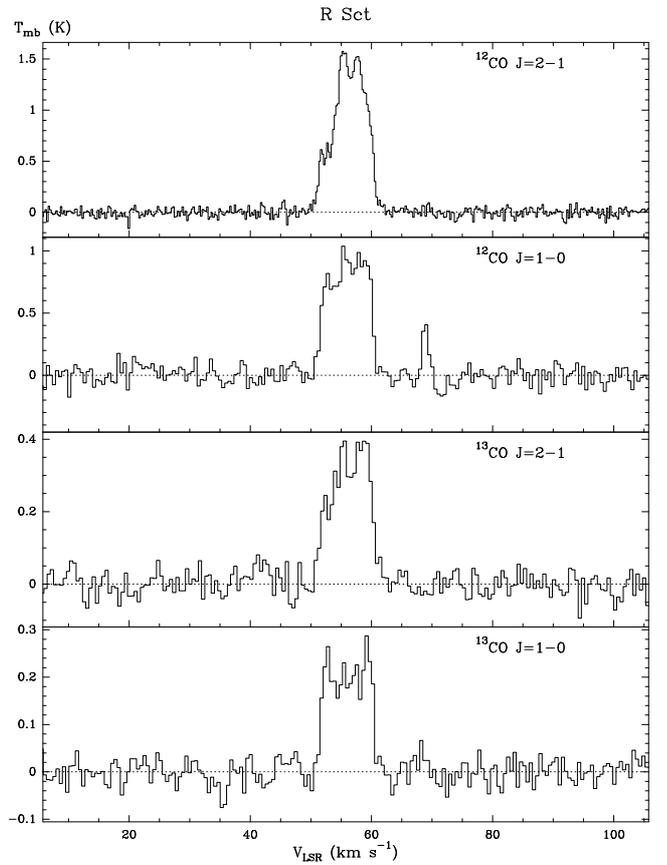}
}}
   \caption{Observed profiles in R Sct. The feature at about 70
       \kms\ LSR in \doce\ \juc\ is very probably interstellar
       contamination.}
              \label{}%
    \end{figure}

   \begin{figure}
   \centering \rotatebox{0}{\resizebox{8.5cm}{!}{ 
\includegraphics{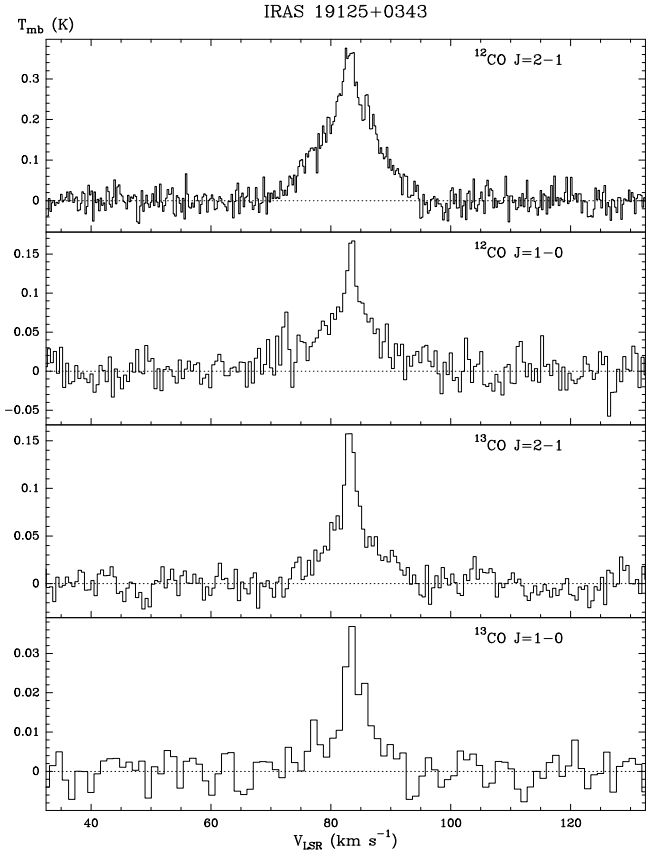}
}}
   \caption{Observed profiles in IRAS 19125+0343. }
              \label{}%
    \end{figure}

   \begin{figure}
   \centering \rotatebox{0}{\resizebox{8.5cm}{!}{ 
\includegraphics{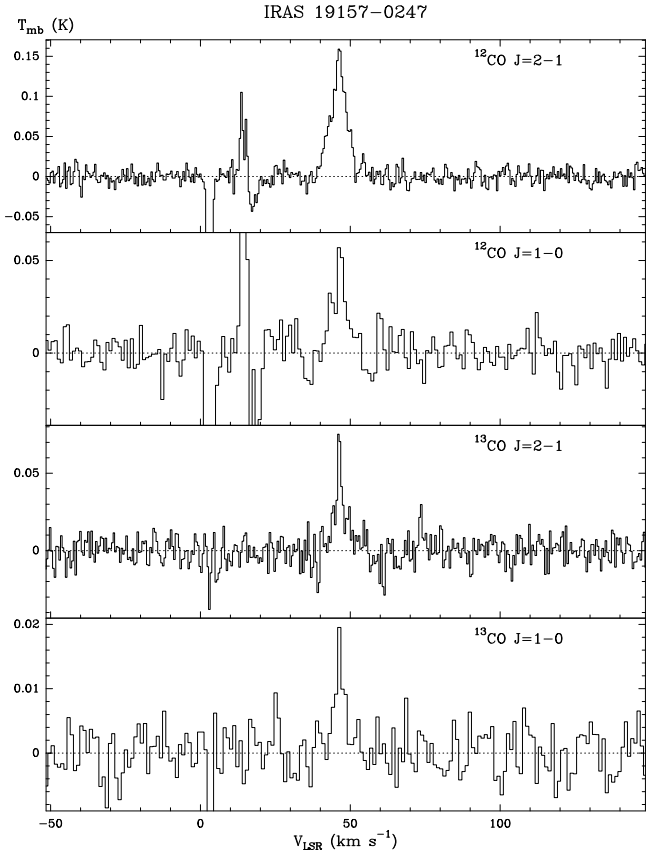}
}}
   \caption{Observed profiles in IRAS 19157-0247. Note the presence of
     interstellar contamination between 0 and 20 \kms\ LSR}
              \label{}%
    \end{figure}

   \begin{figure}
   \centering \rotatebox{0}{\resizebox{8.5cm}{!}{ 
\includegraphics{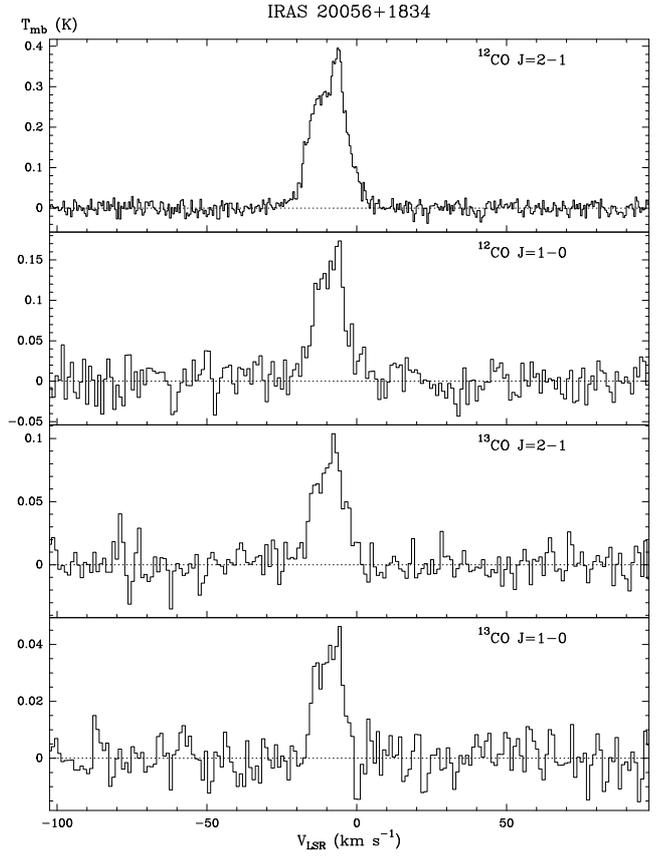}
}}
   \caption{Observed profiles in IRAS 20056+1834. }
              \label{}%
    \end{figure}

\subsection{APEX data}

The data for the southern sources sample have been obtained using the
APEX\footnote{The Atacama Pathfinder EXperiment, APEX, is a
  collaboration between Max Planck Institut f\"ur Radioastronomie
  (MPIfR) at 50\%, Onsala Space Observatory (OSO) at 23\%, and the
  European Southern Observatory (ESO) at 27\% to construct and operate
  a modified ALMA prototype antenna as a single dish on the
  high-altitude site of Llano Chajnantor.}  12\,m telescope, in
April--May 2012. The precipitable water vapor during the run ranged
between 1.0 and 2.5\,mm in general.

We have observed the \doce\ \jdu\ and \jtd\ lines using the SHeFI APEX
Band 1 \citep[][]{vassilev2008} \& Band 2 \citep{risacher2006}
receivers respectively, connected to the Fast Fourier Transform
Spectrometer \citep[XFFTS,][]{klein2012}. The observation was done in the
standard ON--OFF mode using the wobbling secondary. In this mode, the
secondary mirror of the telescope wobbles at the frequency of 1\,Hz
between the ON and OFF positions. At the same time, the telescope nods
such that the OFF positions are alternatively taken at both sides of
the source, always at $\pm$50$''$ in azimuth.

The SHeFI receivers can only be used in one band at a time, so the
observation of the two lines in each object are not simultaneous. For
the \jdu\ line, the XFFTS was configured to work with 13107 channels of
0.099\,\kms\ (76.3\,kHz) of resolution covering a band of about
1300\,\kms. For the \jtd\ transition, the number of channels was 6553
with a resolution of 0.13\,\kms\ (152.6\,kHz), providing a total
bandwidth of 866\,\kms. The data are originally calibrated in terms of
$T_{\rm a}^*$, using the standard procedure of observing hot and cold
loads, and the blank sky.

The data have been inspected and, after removing bad scans, 
time-averaged. After that, baselines of at most degree two have been
removed using the ranges of the spectrum free from line features. The
resolution of the APEX spectra have been downgraded to about 0.5\,\kms
or 1.0\,\kms\ by averaging adjacent channels.  Finally, the data have
been rescaled into \Tmb\ units, using the values for the telescope
efficiencies given by \citet{gusten2006}, see results in Table 4.  We
give more details on the data analysis in Appendix B, where the
resulting APEX spectra are shown, see Figs.\ B.1 to B.10.

Note that in more than one half of the sources, the spectra show
contamination due to emission of interstellar gas in the direction of
the source and of the reference (OFF) position. In some cases this
contamination does not harm our observation, but in others it is so
strong and widespread around the systemic velocity of the source that
we can not conclude on the presence of any circumstellar
contribution. These cases are noted in Table 4.  The presence of such
contamination is not a surprise since many of the selected sources are
located very close to the galactic plane.

\begin{table*}[bthp]
\begin{center}                                          
\caption{Summary of observational results for northern sources (IRAM
  30m telescope). The velocity range is that
  occupied by the full line profile, used in particular to
  calculate the line area. For undetected sources, the velocity range
  is the total analyzed range (centered on the velocity of the source
  deduced from optical data). The uncertainty of the area has been
  calculated taking into account the number of spectral channels
  included in the equivalent line width (given in the last column, see
  text). An asterisk indicates value assumed from data of other
  lines. 
  Uncertain values are indicated by the symbol $\sim$.}

\scriptsize

\begin{tabular}{|ll|ccccccc|l|}
\hline\hline
 & & & & & & & & & \\ 
  Source & transition & velocity range & sp.\ resol.\ & $T_{\rm
    mb}$(peak) & rms & area & rms(area) & eq.\ line width  & comments \\
 & & \kms(LSR) & \kms & K & K & K \kms & K \kms & \kms & \\
\hline
 & & & & & & & & & \\ 
RV Tau & $^{12}$CO 2--1 & 17 $\pm$200 & 0.51 & & 2 10$^{-2}$ & & & &
undetected \\
 & $^{12}$CO 1--0 & 17 $\pm$200 & 0.51 & & 3.5 10$^{-2}$ & & & & \\
 & $^{13}$CO 2--1 & 17 $\pm$100 & 0.53 & & 2.5 10$^{-2}$ & & & & CO emission \\
 & $^{13}$CO 1--0 & 17 $\pm$200 & 0.53 & & 1.5 10$^{-2}$ & $<$6 10$^{-2}$ & & 5$^*$ & \\
\hline
 & & & & & & & & & \\
DY Ori & $^{12}$CO 2--1 & -13 $\pm$5 & 1.02 & 4.5 10$^{-2}$ & 1
10$^{-2}$ & 0.28 & 3 10$^{-2}$ & 6.2 & \\ 
       & $^{12}$CO 1--0 & -13 $\pm$5 & 1.02 & 3.1 10$^{-2}$ & 1
10$^{-2}$ & 0.15 & 2.5 10$^{-2}$ & 5.5 & \\ 
       & $^{13}$CO 2--1 & -13 $\pm$5 & 1.06 & 2.7 10$^{-2}$ & 1
10$^{-2}$ & 0.15 & 2.5 10$^{-2}$ & 5.6 & \\  
         & $^{13}$CO 1--0 & -13 $\pm$5 & 1.06 & $<$1 10$^{-2}$ & 4
10$^{-3}$ & $<$3 10$^{-2}$ & 1.0 10$^{-2}$ & 6$^*$ & \\ 
\hline
 & & & & & & & & & \\
Red Rectangle & $^{12}$CO 2--1 & 0 $\pm$12 & 0.25 & 1.43 & 3.5
10$^{-2}$ & 7.00 & 4 10$^{-2}$ & 4.9 &  \\
 & $^{12}$CO 1--0 & 0 $\pm$12 & 0.51 & 0.34 & 2.5 10$^{-2}$ & 1.53 & 4
10$^{-2}$ & 4.5 &  \\
 & $^{13}$CO 2--1 & 0 $\pm$12 & 0.27 & 0.61 & 1.5 10$^{-2}$ & 2.43 & 2 10$^{-2}$ & 4.0 & \\
 & $^{13}$CO 1--0 & 0 $\pm$12 & 0.53 & 0.13 & 9.5 10$^{-3}$ & 0.47 & 1.5 10$^{-2}$ & 3.65 & \\
\hline
 & & & & & & & & & \\
U Mon & $^{12}$CO 2--1 & 9 $\pm$200 & 0.51 &  & 1.5
10$^{-2}$ &  &  &  & undetected \\
 & $^{12}$CO 1--0 & 9 $\pm$200 & 0.51 &  & 2 10$^{-2}$ &  &  & & \\
 & $^{13}$CO 2--1 & 9 $\pm$100 & 0.53 &  & 2 10$^{-2}$ &
 &  &  & CO emission \\
 & $^{13}$CO 1--0 & 9 $\pm$200 & 0.53 &  & 1 10$^{-2}$ & $<$5
10$^{-2}$ &  & 5$^*$ &  \\
\hline
 & & & & & & & & & \\
AI CMi & $^{12}$CO 2--1 & 29 $\pm$10 & 0.25 & 0.50 & 3.5 10$^{-2}$ &
3.43 & 4.5 10$^{-2}$ & 6.9 & \\ 
       & $^{12}$CO 1--0 & 29 $\pm$10 & 0.51 & 0.16 & 2 10$^{-2}$ &
0.95 & 3.5 10$^{-2}$ & 5.9 & \\ 
       & $^{13}$CO 2--1 & 29 $\pm$10 & 0.53 & 0.18 & 1.5 10$^{-2}$ &
1.4 & 3 10$^{-2}$ & 7.7 & \\ 
       & $^{13}$CO 1--0 & 29 $\pm$10 & 1.06 & 3.5 10$^{-2}$ & 5
10$^{-3}$ & 0.31 & 1.5 10$^{-2}$ & 8.8 & \\ 
\hline
 & & & & & & & & & \\ 
HR 4049 & $^{12}$CO 2--1 & -45 $\pm$6 & 0.51 & 0.11 & 2 10$^{-2}$ & 0.50 & 3 10$^{-2}$ & 4.5 & \\
 & $^{12}$CO 1--0 & -45 $\pm$6 & 1.02 & $\sim$4 10$^{-2}$ & 2 10$^{-2}$ & $\sim$0.1 & 4 10$^{-2}$ & 3.3 & \\
 & $^{13}$CO 2--1 & -45 $\pm$6 & 1.06 & 3.0 10$^{-2}$ & 1 10$^{-2}$ & 0.15 & 2.5 10$^{-2}$ & 5.7 & \\
 & $^{13}$CO 1--0 & -45 $\pm$6 & 1.06 & $\sim$1.6 10$^{-2}$ & 6 10$^{-3}$ & 5.4 10$^{-2}$ & 1.5 10$^{-2}$ & 4.9 & \\
\hline
 & & & & & & & & & \\
89 Her & $^{12}$CO 2--1 & -8 $\pm$14 & 0.25 & 1.78 & 2.5 10$^{-2}$ &
9.06 & 2.5 10$^{-2}$ & 5.1 &  \\
 & $^{12}$CO 1--0 & -8 $\pm$14 & 0.51 & 0.81 & 2.5 10$^{-2}$ & 5.01 & 4
10$^{-2}$ & 6.2 &  \\
 & $^{13}$CO 2--1 & -8 $\pm$14 & 0.27 & 0.80 & 2.5 10$^{-2}$ & 2.34 & 2 10$^{-2}$ & 2.9 & \\
 & $^{13}$CO 1--0 & -8 $\pm$14 & 0.53 & 0.23 & 1 10$^{-2}$ & 0.97 & 1.5 10$^{-2}$ & 4.2 & \\
\hline
 & & & & & & & & & \\
IRAS 18123+0511 & $^{12}$CO 2--1 & 95 $\pm$25 & 0.51 & 0.12 & 1.5
10$^{-2}$ & 2.16 & 4 10$^{-2}$ & 18.0 & \\ 
         & $^{12}$CO 1--0 & 95 $\pm$25 & 0.51 & 4.3 10$^{-2}$ & 1
10$^{-2}$ & 0.71 & 3.5 10$^{-2}$ & 16.5 & \\ 
         & $^{13}$CO 2--1 & 95 $\pm$25 & 0.53 & 4.8 10$^{-2}$ & 6.5
10$^{-3}$ & 0.53 & 1.5 10$^{-2}$ & 11.1 & \\ 
         & $^{13}$CO 1--0 & 95 $\pm$25 & 0.53 & 1.5 10$^{-2}$ & 4
10$^{-3}$ & 0.14 & 9 10$^{-3}$ & 9.3 & \\ 
\hline
 & & & & & & & & & \\
AC Her & $^{12}$CO 2--1 & -10 $\pm$5 & 0.25 & 0.17 & 2 10$^{-2}$ &
0.44 & 1.5 10$^{-2}$ & 2.6 & \\ 
       & $^{12}$CO 1--0 & -10 $\pm$5 & 0.51 & 3.2 10$^{-2}$ & 1
10$^{-2}$ & 5.5 10$^{-2}$ & 1 10$^{-2}$ & 1.8 & \\ 
       & $^{13}$CO 2--1 & -10 $\pm$5 & 0.27 & 7.2 10$^{-2}$ & 9.5
10$^{-3}$ & 0.21 & 9 10$^{-3}$ & 2.9 & \\ 
       & $^{13}$CO 1--0 & -10 $\pm$5 & 1.06 & 8.0 10$^{-3}$ & 2.5
10$^{-3}$ & 2.5 10$^{-2}$ & 4.5 10$^{-3}$ & 3.0 & \\ 
\hline
 & & & & & & & & & \\
R Sct & $^{12}$CO 2--1 & 56 $\pm$8 & 0.25 & 1.54 & 4 10$^{-2}$ & 10.2
& 5 10$^{-2}$ & 6.6 & composite \\ 
      & $^{12}$CO 1--0 & 56 $\pm$8 & 0.51 & 0.98 & 6.5 10$^{-2}$ & 7.40
& 1 10$^{-1}$ & 7.6 & \\ 
      & $^{13}$CO 2--1 & 56 $\pm$8 & 0.27 & 0.39 & 5 10$^{-2}$ & 2.84 & 7
10$^{-2}$ & 7.4 &  profiles \\ 
      & $^{13}$CO 1--0 & 56 $\pm$8 & 0.53 & 0.26 & 2.5 10$^{-2}$ & 1.82
& 5 10$^{-2}$ & 7.0 & \\ 
\hline
 & & & & & & & & & \\
IRAS 19125+0343 & $^{12}$CO 2--1 & 83 $\pm$15 & 0.25 & 0.35 & 2.5
10$^{-2}$ & 3.31 & 3.5 10$^{-2}$ & 9.5 & \\ 
                & $^{12}$CO 1--0 & 83 $\pm$15 & 0.51 & 0.15 & 1.5
10$^{-2}$ & 1.31 & 3.5 10$^{-2}$ & 8.7 & \\ 
                & $^{13}$CO 2--1 & 83 $\pm$15 & 0.27 & 0.15 & 1.5
10$^{-2}$ & 0.89 & 2 10$^{-2}$ & 5.9 & \\ 
                & $^{13}$CO 1--0 & 83 $\pm$15 & 0.53 & 3.8 10$^{-2}$ &
6.5 10$^{-3}$ & 0.21 & 1 10$^{-2}$ & 5.5 & \\ 
\hline
 & & & & & & & & & \\
IRAS 19157-0247 & $^{12}$CO 2--1 & 46 $\pm$7 & 0.51 & 0.15 & 9 10$^{-3}$ &
0.99 & 1.5 10$^{-2}$ & 6.6 & \\ 
           & $^{12}$CO 1--0 & 46 $\pm$7 & 1.02 & 5.1 10$^{-2}$ & 8
10$^{-3}$ & 0.34 & 2.1 10$^{-2}$ & 6.5 & \\ 
           & $^{13}$CO 2--1 & 46 $\pm$7 & 0.53 & 7.0 10$^{-2}$ & 8.5 
10$^{-3}$ & 0.25 & 1 10$^{-2}$ & 3.6 & \\
           & $^{13}$CO 1--0 & 46 $\pm$7 & 1.06 & 1.6 10$^{-2}$ & 3.5
10$^{-3}$ & 6.0 10$^{-2}$ & 7 10$^{-3}$ & 3.8 & \\ 
\hline
 & & & & & & & & & \\
IRAS 20056+1834 & $^{12}$CO 2--1 & -9 $\pm$20 & 0.25 & 0.39 & 2
10$^{-2}$ & 4.91 & 3.5 10$^{-2}$ & 12.6 & composite \\
 & $^{12}$CO 1--0 & -9 $\pm$20 & 0.51 & 1.8 10$^{-2}$ & 2.5 10$^{-2}$ & 1.85 & 6 10$^{-2}$ & 10.2 & \\
 & $^{13}$CO 2--1 & -9 $\pm$20 & 0.53 & 9.5 10$^{-2}$ & 1.5 10$^{-2}$ &
1.08 & 4 10$^{-2}$ & 11.4 &  profiles \\
 & $^{13}$CO 1--0 & -9 $\pm$20 & 0.53 & 4.5 10$^{-2}$ & 1 10$^{-2}$ & 0.42 & 2 10$^{-2}$ & 9.3 & \\
\hline
 & & & & & & & & & \\
R Sge & $^{12}$CO 2--1 & 28 $\pm$200 & 0.51 & & 1 10$^{-2}$ & & & &
undetected \\
 & $^{12}$CO 1--0 & 28 $\pm$200 & 0.51 & & 2.5 10$^{-2}$ & & & & \\
 & $^{13}$CO 2--1 & 28 $\pm$ 100 & 0.53 & & 1.5 10$^{-2}$ & & & & CO
emission \\
 & $^{13}$CO 1--0 & 28 $\pm$200 & 0.53 & & 1 10$^{-2}$ & $<$5
10$^{-2}$ & & 5$^*$ & \\ 

\hline\hline
\end{tabular}
\end{center}
\end{table*}

\begin{table*}
\begin{center}                                          
\caption{Summary of APEX observations of southern sources. The velocity
  range is that occupied by the full line profile, used in particular
  to calculate the line area. For undetected sources, the velocity
  range is the total analyzed range (centered on the velocity of the
  source deduced from optical data). The uncertainty of the area has
  been calculated taking into account the number of spectral channels
  included in the equivalent line width (given in the last column, see
  text). An asterisk indicates value assumed from data of other
  lines. Uncertain values are indicated by the symbol $\sim$.}

\scriptsize
\begin{tabular}{|ll|ccccccc|l|}
\hline\hline
 & & & & & & & & & \\ 
  Source & transition & velocity range & sp.\ resol.\ & $T_{\rm
    mb}$(peak) & rms & area & rms(area) & eq.\ line width  & comments \\
 & & \kms(LSR) & \kms & K & K & K \kms & K \kms & \kms & \\
\hline
 & & & & & & & & & \\ 
AR Pup & $^{12}$CO 3--2 & 29 $\pm$200 & 0.53 & & 0.02 & & & &
ISM narrow line  \\
& $^{12}$CO 2--1 & 29 $\pm$200 & 0.6 & & 0.06 & & & & ISM narrow line \\
\hline
 & & & & & & & & & \\
IRAS 08544-4431 & $^{12}$CO 3--2 & 45 $\pm$12 & 0.53 & 0.56 & 0.02 &
4.22 & 0.035 & 7.5 & \\
 & $^{12}$CO 2--1 & & & 0.45 & & 3 & & & SEST data (Maas et al.) \\
\hline
 & & & & & & & & & \\
IW Car & $^{12}$CO 3--2 & --28 $\pm$9 & 0.53 & 0.22 & 0.03 & 0.71 & 0.035 &
3.2 & \\
 & $^{12}$CO 2--1 & --28 $\pm$9 & 0.6 & 0.14 & 0.015 & 0.54 & 0.025 &
3.9 & \\ 
\hline
 & & & & & & & & & \\
IRAS 10174-5704 & $^{12}$CO 3--2 & 3 $\pm$200 & 0.53 & & 0.04 & & & &
strong IS cont.\ \\
 & $^{12}$CO 2--1 & 3 $\pm$200 & 0.6 & & 0.02 & & & & strong IS
cont.\ \\ 
\hline
 & & & & & & & & & \\
IRAS 10456-5712 & $^{12}$CO 3--2 & --9 $\pm$200 & 0.53 & & 0.03 & & &
& strong IS cont.\ \\ 
 & $^{12}$CO 2--1 & --9 $\pm$200 & 0.6 & & 0.015 & & & & strong IS
cont.\ \\  
\hline
 & & & & & & & & & \\
HD 95767 & $^{12}$CO 3--2 & --32 $\pm$50 & 1.1 & 0.05 & 0.013 &
0.26 & 0.03 & 5.2 & \\
 & $^{12}$CO 2--1 & --32 $\pm$50 & 1 & $\sim$0.03 & 0.01 & $\sim$0.06 &
0.02 & 5.2$^*$ & \\
\hline
 & & & & & & & & & \\
RU Cen & $^{12}$CO 3--2 & --35 $\pm$200 & 0.53 & & 0.02 & & & & undetected CO
em.\ \\
 & $^{12}$CO 2--1 & --35 $\pm$200 & 0.6 & & 0.015 & & & &  undetected CO
em. \\
\hline
 & & & & & & & & & \\
HD 108015 &  $^{12}$CO 2--1 & --2 $\pm$6 & 0.6 & 0.11 & 0.02 & 0.72 &
0.04 & 6.5 & \\
\hline
 & & & & & & & & & \\
IRAS 15469-5311 & $^{12}$CO 3--2 & --14 $\pm$200 & 0.53 & & 0.02
& & & & strong IS cont.\ \\  
& $^{12}$CO 2--1 & --14 $\pm$200 & 0.6 & & 0.02 & & & & strong IS
cont.\ \\  
\hline
 & & & & & & & & & \\
IRAS 15556-5444 & $^{12}$CO 3--2 & 0 $\pm$200 & 0.53 & & 0.02
& & & & IS contamination \\
 & $^{12}$CO 2--1 & 0 $\pm$200 & 0.6 & & 0.015
& & & & IS contamination \\

\hline\hline
\end{tabular}
\end{center}
\end{table*}

\section{Rotating disks in the observed sources}

We have observed 24 post-AGB stars in which the presence of inner, very
compact rotating disks was suspected from the presence of a significant
NIR excess (other properties of the sources are mentioned in Sects.\ 1,
2). Fourteen sources were observed with the 30m IRAM telescope in
\doce\ and \trece\ \jdu\ and \juc\ transitions, and ten southern
objects were observed with APEX in \doce\ \jtd\ and \jdu. The detection
rate is very high: of the 14 sources observed with the 30m telescope,
we detected CO emission in 11.
We also detected disk emission from 4 southern post-AGB stars. However,
because of the lower S/N ratio attained with the APEX telescope and the
presence of significant contamination by interstellar emission in many
sources (Sect.\ 3.2, Appendix B), any estimate of a detection rate
comparable to that deduced for the northern sources is impossible.

Only four of these sources were previously detected in CO lines, the
Red Rectangle, 89 Her, R Sct, and IRAS\,08544-4431, plus a tentative
detection of AC Her, with line parameters similar to those found here
but in all cases with significantly less complete and accurate
profiles.
See previous molecular line data in Bujarrabal et al.\ (2005, 2007),
Jura et al.\ (1995), Alcolea \& Bujarrabal (1995), Maas et al.\ (2003),
and references therein.

The Red Rectangle, the best studied source in our sample, shows the
characteristic profile of rotating disks, narrow and with a central
peak or two nearby central peaks, see Fig.\ 2. This kind of line shape
is often observed in disks around young stars and has been very well
studied. It is widely accepted, both from observational and theoretical
grounds, that these profiles are a good probe of the presence of disks
in rotation (Guilloteau \& Dutrey 1998, Guilloteau et al.\ 2013,
Bujarrabal et al.\ 2013, etc). The CO line profiles observed in evolved
nebulae of other classes are very different: in AGB stars the profiles
are wide, often wider than 20 \kms, with two horns or parabolic, and in
other {\em standard} PPNe they are even broader, sometimes as wide as
100 \kms, and show a core+wings structure.  Interferometric maps of the
Red Rectangle (Bujarrabal et al.\ 2005) conclusively confirmed that a
rotating disk is the dominant nebular component in mm-wave line
emission. The maps show that the CO-rich disk is extended (about 6$''$
in diameter, equivalent to about 6 10$^{16}$ cm, and \lsim\ 1$''$ in
width), obviously in rotation, and placed in the equator of the optical
nebula.  High-$J$ Herschel/HIFI observations show an excess in the line
wings that seem impossible to explain from disk emission alone, this
excess is also present though less clearly in low-$J$ transitions, see
data and detailed models in Bujarrabal et al.\ (2013). According to
these authors, this excess comes probably from a relatively diffuse
bipolar component expanding at about 10 \kms. This feature is
conspicuous in the \doce\ lines but very weak in \trece\ lines. This
properties indicates that the CO emission from the bipolar outflow is
probably optically thin and that the rare-isotope lines would only come
from the disk. These suggestions have been spectacularly confirmed by
very recent ALMA maps of \doce\ and \trece\ lines (paper in
preparation). Therefore, emission from a bipolar outflow significantly
contributes to the wings of the profiles shown here, mostly those of
\doce.

89 Her has been also mapped in CO lines (Bujarrabal et al.\ 2007). An
expanding double-bubble is responsible for most emission at velocities
separated more than 3--4 \kms\ from the line center. There is also a
practically unresolved component in the center that is very probably a
rotating disk, in view of the low velocities involved and the very
similar single-dish CO profiles shown by 89 Her and the Red Rectangle
(Figs.\ 2 and 5).

Both in the Red Rectangle and 89 Her, the mass of the outflow is
smaller than that contained in the disk, which is the dominant nebular
component. In any case, the total mass of the rotating or expanding gas
around these stars remains moderate, \lsim\ 10$^{-2}$ \ms. See Sect.\ 5
for more details.

Of the rest of the northern sources detected here, six (DY Ori,
HR\,4049, IRAS\,18123+0511, AC Her, IRAS\,19125+0343, and
IRAS\,19157-0247) show narrow profiles very similar to those of the Red
Rectangle and 89 Her, and two (R Sct and IRAS\,20056+1834) show
composite profiles including a narrow component similar to the main one
found in the other sources. In AI CMi, the profiles are narrow but less
sharp than in the Red Rectangle or 89 Her. AI CMi shows OH
emission with a two-horn profile similar to those typical of AGB stars,
except for the very small expansion velocity (and the high polarization
degree in 1612 MHz emission, typical of post-AGB OH emitters, Wolak et
al.\ 2012). OH emission expands between 24 and 33 \kms\ $LSR$, as our
CO lines. The narrow profiles and low-intensity wings of our CO lines
in AI CMi strongly suggest emission from a rotating disk, perhaps
superimposed on emission from an expanding component. The line wings of
IRAS\,18123+0511 and IRAS\,19125+0343 are conspicuous, around a
prominent central component. These wings are particularly intense in
the \doce\ \jdu\ line, the most opaque of our lines, but much less
evident in \trece\ emission. In these two sources, we find therefore a
dominant disk emission plus, probably, a significant contribution from
an expanding component.
The profiles of R Sct are quite complex, the central spike, perhaps due
to disk emission, just represents a fraction of the total intensity.
IRAS\,20056+1834 also shows complex profiles with a narrow peak. The
uncertain stellar velocity of IRAS\,20056+1834 deduced from studies in
the optical, $\sim$ 0 \kms, is close to the velocity of the narrow peak
but more positive than the centroid of the line-wings, which appear
clearly blue-shifted. We note that in these two sources, and contrary
to the case of the objects discussed above, the outflow feature does
not become relatively weaker for low-opacity lines.

We therefore conclude that in 9 sources (the Red Rectangle, 89 Her, DY
Ori, HR\,4049, IRAS\,18123+0511, AC Her, IRAS\,19125+0343, IRAS\,19157-0247,
and probably AI CMi), out of a total of 14 sources in our sample of
sources accurately observed with the 30-telescope, the molecule-rich
nebula is dominated by an extended rotating disk.
The Red Rectangle and 89 Her show intense line wings that are known,
from high-resolution maps, to come from low-mass bipolar outflows with
velocities of 5 -- 10 \kms.  Two of the other objects in this group,
IRAS\,18123+0511 and IRAS\,19125+0343, also show intense line wings, at
velocities of $\pm$ 5 -- 10 \kms\ from the center, that very probably
come from a low-mass bipolar outflow similar to those of the Red
Rectangle and 89 Her. AI CMi could also show emission from expanding
gas, in view of the relatively wide lines found in this source. Such
slow bipolar outflows would then be associated to the presence of
equatorial disks in rotation. We recall that bipolar outflows are found
in many {\em standard} PPNe (see Sect.\ 1), though showing much
higher masses and velocities than those found here (Sect.\ 5).

Two other objects, R Sct and IRAS\,20056+1834, show complex lines with
a narrow component (about 5 \kms\ wide), different than those discussed
in the previous paragraph. We suggest that this narrow feature comes
from a rotating disk that, in these nebulae, only contains a fraction
of the total mass. The rest of the CO emission detected in these two
objects (whose line are in total 10 -- 15 \kms\ wide) would probably
come from low-velocity outflows.

Of the ten southern sources observed with APEX, four clearly show
disk-like CO profiles: IRAS\,08544-4431, IW Car, HD\,95767, and
HD\,108015. The intense \doce\ \jtd\ line of IRAS\,08544-4431
(Fig.\ B.2) presents wide wings suggesting a contribution of an
expanding component. The \doce\ \jtd\ and \juc\ lines observed by Maas
et al.\ (2003) in this source show similar characteristics. As
mentioned, the lower quality of the observations of these southern
sources prevents an analysis of the data as deep as for the northern
sources.

\section{Estimates of the rotating disk mass in post-AGB objects}

\subsection{Basic relations between the total emitting gas mass and the
 observed line intensity} 

From our observational results, it is possible to systematically derive
estimates of the total mass of molecule-rich gas in the studied
nebula. If we assume optically thin emission, that the emitting region
is much smaller than the telescope resolution, and that some
characteristic excitation state can represent the whole emission of the
disk, the total mass is just proportional to the velocity-integrated
main-beam temperature. 

Optically thin \trece\ lines are expected, in view of the significantly
low \trece/\doce\ line ratio, mostly for the \juc\ transition (but the
\trece/\doce\ intensity ratio is usually found to be higher than the
expected abundance ratios, suggesting optically thick \doce\ emission,
mostly in the line core).  Our detailed calculations for the case of
the Red Rectangle (Bujarrabal et al.\ 2013) also indicate optically
thin \trece\ lines. Note that \trece\ lines are also preferable to
estimate the mass of the rotating disks because in them the
contribution of the eventual expanding components (found in the well
studied cases, Sects.\ 1, 4) is much smaller, probably negligible in
most cases. Note that the \juc\ transition is a priori expected to be
less opaque than the \jdu\ one, because a) both transitions are easily
excited (the upper level of \trece\ $J$=2--1 is just at 16 K from the
ground), b) the low Einstein A-coefficients, $A$(1--0) = 6.3 10$^{-8}$
s$^{-1}$, guarantee an easy thermalization of the low-$J$ lines (see
calculations in e.g.\ Bujarrabal et al.\ 2005, 2013), and c) the
opacity of the rotational line is roughly proportional to $J^2$ under
the above conditions.

The assumption of an emitting region smaller than the telescope beam is
very probably satisfied in our case, mostly for the \juc\ lines, since
the size of the CO images obtained in 89 Her and the Red Rectangle
(which are the strongest emitters among our sources and the only ones
for which good maps have been obtained, Sects.\ 1, 4), are $\sim$
6$''$, and the beam half-power total widths are about 12$''$ at 1mm
wavelength and about 23$''$ at 3mm. We have also seen in Sect.\ 4 that
the \trece\ emission of our sources is more compact, since the emission
of the disk is not contaminated by components in expansion.

Under the above conditions, the observed main-beam temperature of a
given transition is given by:
\begin{equation}
T_{\rm mb}(V_{LSR})  ~=~ \frac{h \nu}{k}~ [\frac{1}{e^{h
      \nu/T_{\rm ex}} - 1} - \frac{1}{e^{h\nu/T_{\rm BG}}-1}] ~\tau ~\phi(V_{LSR})
  ~\frac{\Omega_{\rm s}}{\Omega_{\rm mb}} .
\end{equation}
Where $\Omega_{\rm mb}$ and $\Omega_{\rm s}$ are respectively the solid
angles subtended by the telescope beam and the source (we assume
$\Omega_{\rm s}$ $<$ $\Omega_{\rm mb}$), $T_{\rm ex}$ and $T_{\rm BG}$
are the excitation temperature of the transition and the background
temperature (we assume $T_{\rm ex}$ $\sim$ $T_{\rm k}$, the kinetic
temperature, and $T_{\rm BG}$ = 2.7 K), $\nu$ is the frequency of the
transition, and $\phi(V_{LSR})$ is the normalized line profile in terms
of Doppler-shifted velocity ($LSR$ frame); the other symbols have their
usual meaning.

The opacity is   
\begin{equation}
 \tau  = \frac{c^3}{8\pi\nu^3} ~A_{u,l} ~g_u ~(x_l-x_u) ~n_{\rm tot} ~X({\rm
   CO}) ~L ~~.
\end{equation}
Where the subindexes $l,u$ represent respectively the lower and upper
level of the transition, $g$ is the statistical weight of a level, $x$
is the relative population per magnetic sublevel (being $n$ the
relative level population, i.e.\ $x$ = $n/g$), $n_{\rm tot}$ is the
total number density of particles, $X$(CO) is the abundance of
\doce\ or \trece\ relative to the total number of particles, and $L$ is
the typical length of the source along the line of sight (perpendicular
to the plane of the sky). This expression is perhaps not the most usual
one of the optical depth, it is used because the profile is given in
terms of velocity and not of frequency.

Converting the source solid angle (in arcsec units) into source surface
(cm$^2$), the velocity integrated main-beam temperature becomes 
\begin{equation}
\int T_{\rm mb} ~{\rm d}V ~\propto ~ [\frac{1}{e^{h \nu/T_{\rm ex}} -
    1} - \frac{1}{e^{h\nu/T_{\rm BG}}-1}] ~(x_l-x_u) ~X ~M_{\rm mol} ~D^2 ,
\end{equation}
where $M_{\rm mol}$ is the total mass of the molecule-rich emitting gas
(i.e.\ the characteristic total density multiplied by the source
volume) and $D$ is the distance to the object. So, we can estimate the
total mass from the observed profile area for optically thin,
unextended emission.  The proportionality factor is straightforward
from eqs.\ 1 and 2.

Note that in eq.\ 3 
$$
\frac{1}{e^{h \nu/T_{\rm ex}}-1} = \frac{x_u}{x_l-x_u}
$$
and 
$$
x_{l,u} = e^{-E_{l,u}/T_{\rm exc}}/F(T_{\rm rot}) ~~,
$$ where $E$ is the energy in temperature units of the considered
levels (in our case, $J$ = 0, 1, or 2) and $F$ is the partition
function for a representative {\em rotational} temperature, $T_{\rm
  rot}$, that describes the population of the whole rotational
ladder. For an actual value of the partition function, $T_{\rm rot}$ is
defined as the temperature that gives that value assuming
thermalization of the level populations. For a simple rotational ladder
and in the limit of $T_{\rm rot}$ much higher than the separation of
the low levels, $F(T_{\rm rot})$ = $T_{\rm rot}$/$B_{\rm rot}$, where
$B_{\rm rot}$ is the rotational constant of the molecule (for \trece,
$B_{\rm rot}$ = 2.6 K). In our calculations, we always used the exact
expression:
$$
F(T_{\rm rot}) = \sum_J g_J e^{-E_j/T_{\rm rot}}~~.
$$ We cannot assume in our case (with relatively high values of $T_{\rm
  k}$) that $T_{\rm rot}$ $\sim$ $T_{\rm k}$ ($\sim$ $T_{\rm ex}$),
since the low-$J$ levels joined by the transitions we are dealing with
are easily populated, but the population of high-$J$ levels is probably
strongly subthermal, so $x_J/x_0$ $<$ $e^{-E_J/T_{\rm k}}$ and
$T_{\rm rot}$ $<$ $T_{\rm k}$ $\sim$ $T_{\rm ex}$. We will see later
estimates of the values we can expect for $T_{\rm rot}$.  Taking into
account that in our case 
$T_{\rm BG}$ $\ll$ $T_{\rm ex}$ and $E_u$ $\ll$ $T_{\rm ex}$, the
dependence of the total mass on $T_{\rm ex}$ practically vanishes and
the dependence of the total mass estimate on the level excitation is
basically given by the rotational temperature.

\subsection{Estimates of the CO rotational temperature in post-AGB
  rotating disks}

The temperature of the molecular gas in the Red Rectangle was first
estimated by Bujarrabal et al.\ (2005), from maps of CO mm-wave lines,
obtaining values between about 70 and 30 K. However, those low-$J$
transitions are not good tools to estimate the excitation of such a
warm gas. Later, Herschel/HIFI observations of the \jsc, \jdn,
\jdq\ transitions were used to deduce that the gas temperature in the
disk of the Red Rectangle must be about two times larger
(Bujarrabal et al.\ 2012, 2013).

The temperature of the molecular (compact) disk of 89 Her was deduced
to be \gsim\ 60 K from mm-wave maps (Bujarrabal et al.\ 2007). Herschel
observations of 89 Her, R Sct, and IRAS\,19125+0511 yield high-$J$
transitions that are weaker by at least a factor 3 than expected from
the low-$J$ line ratios found here and the Herschel/HIFI data of the
Red Rectangle (Teyssier et al., in preparation; Bujarrabal et
al.\ 2013). This result strongly suggests that the temperatures are
clearly lower in these sources than in the Red Rectangle. Comparison of
the Herschel data of 89 Her, R Sct, and IRAS\,19125+0511 with
theoretical line ratios by Bujarrabal et al.\ (2012), calculated under
quite general conditions, indicates temperatures lower than 100
K. Therefore, in 89 Her we can expect 60 K \lsim\ $T_{\rm k}$
\lsim\ 100 K, and similar relations probably hold also for R Sct and
IRAS\,19125+0511.

For the Red Rectangle, typical densities of about 10$^4$--10$^5$
cm$^{-3}$ were deduced (Bujarrabal et al.\ 2013).  The densities
are probably higher in 89 Her, which is more intense than the Red
Rectangle although placed at a larger distance, but most of the other
sources observed here are significantly weaker.

For these values of the temperature and of the density, we can expect
that the low-$J$ transitions are thermalized, but not for $J$ \gsim\ 5,
which show significantly higher Einstein coefficients ($A$ is roughly
proportional to $J^3)$. The populations of the these high-$J$ levels,
which significantly contribute to the partition function for the
relatively high temperatures we are dealing with, cannot be
neglected. Therefore, statistical equilibrium calculations of the level
populations are necessary to estimate the values of the the equivalent
$T_{\rm rot}$ that must be used to calculate the partition function.

We have performed calculations of $T_{\rm rot}$ for a variety of
densities and temperatures in the optically thin limit. We recall that
these values represent the temperature that the gas should have in the
case of level thermalization yielding the same partition function as
that actually given by the calculations. In the optically thin limit,
the details of the radiative transfer treatment are not relevant to the
estimate of the level population (which are in fact given by a simple
non-iterative calculation). We used collisional rates from the
$\lambda$-database (Sch\"oier et al.\ 2005, Yang et al.\ 2010). The
presence of continuum sources has no significant effect in the level
population for the case we are describing (Bujarrabal et al.\ 2013;
Santander-Garc\'{\i}a et al.\ 2013, in preparation). In our simple
case, $T_{\rm rot}$ depends only on the density and the temperature.

   \begin{figure}
   \centering \rotatebox{0}{\resizebox{8.5cm}{!}{ 
\includegraphics{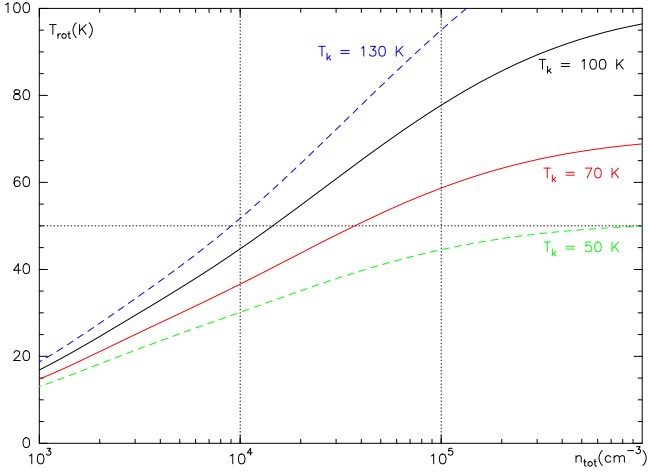}
}}
   \caption{Estimates of the equivalent rotational temperature describing
     the partition function for a variety of densities and
     temperatures. The most probable ranges are of about
     10$^4$ -- 10$^5$ cm$^{-3}$ for the densities (dashed vertical
     lines) and of about 70 -- 100 K for the temperatures (continuous
     lines). }
              \label{}%
    \end{figure}

Our calculations are shown in Fig.\ 12. For the expected ranges of kinetic
temperatures, between approximately 70 K and 100 K, and of total densities,
between approximately 10$^4$ -- 10$^5$ cm$^{-3}$, we expect a
rotational temperature $T_{\rm rot}$ $\sim$ 50 K. The uncertainty of
this value is less than a factor two, which results in a similar
uncertainty for the partition function. We will accordingly adopt
$T_{\rm rot}$ = 50 K in our calculations of the disk mass from the
observed line intensities. 

When \trece\ \juc\ cannot be used (e.g.\ because of very weak
emission), our formulae can be applied to data of other lines. In that
case, we must note that we can underestimate the mass because, as
mentioned, the \doce\ lines (and perhaps \trece\ \jdu) are probably
optically thick; moreover, the \jdu\ lines are probably not less
  extended, because of the moderate excitation required and their
  higher opacities, and so the smaller \jdu\ beam (12--13$''$) can miss
  some emission. Remember that, even if the half-power beam width is
larger than the source size, the telescope gain decreases when it is
not much larger than the source. We also recall (Sect. 5.1) that
\jdu\ is in our cases more opaque than \juc.
In fact, in all objects in which the exercise can be done, the
estimates from \trece\ \juc\ are larger than those from the other
lines, proving that these later computations really underestimate the
actual mass value.

\subsection{The mass of the disks rotating around post-AGB stars}

We use eq.\ 3 (Sect. 5.1) to estimate the mass of the
rotating disks we have found in our sample of post-AGB stars. We will
assume: $T_{\rm rot}$ = 50 K (Sect.\ 5.2), $X$(\trece) = 2 10$^{-5}$
(see below), an average particle mass of 3 10$^{-24}$ g, and $T_{\rm
  ex}$ = $T_{\rm rot}$ (which has little impact on the final values,
Sect.\ 5.2). We use as far as possible the \trece\ \juc\ data, which
are the best for our purposes, as discussed above (Sects.\ 5.1,
5.2). The distances adopted for each object are given in Tables 1 and
2, and the telescope beam widths are given in Sect.\ 3.

\begin{table}[bthp]
\begin{center}                                          
\caption{Values of the nebular mass derived from our data. Distance
  values are discussed in Sect.\ 2. 
For 89 Her, R Sct, and IRAS\,20056+1834, we have calculated the total
nebular mass and also the mass derived from just the central line peak,
expected to better represent the disk emission. See discussion on these
cases and other uncertain values in Sect.\ 5.3 and Appendix B.}

\scriptsize
\vspace{-.1cm}
\begin{tabular}{|l|c|cc|l|}
\hline
Source &   mass & \multicolumn{2}{|c|}{typical size} & comments \\
       & \ms &  $''$ & cm & \\
\hline
 & & & &  \\
RV Tau          & $<$ 8 10$^{-3}$ & $<$0.5 & $<$1.3 10$^{16}$ &  \\
DY Ori          & 2 10$^{-3}$ & 0.37 & 1.1 10$^{16}$  & \\
Red Rectangle   & 6.6 10$^{-3}$ & & & total neb.\ \\
Red Rectangle   & 6 10$^{-3}$ &  2 & 2.3 10$^{16}$  & disk  \\
U Mon           & $<$ 9 10$^{-4}$ & $<$0.4 & $<$ 5 10$^{15}$ & \\
AI CMi          & 1.9 10$^{-2}$ & 1.2 & 2.7 10$^{16}$ & prob.\ exp.\ comp. \\
HR 4049         & 6.3 10$^{-4}$ & 0.6 & 6 10$^{15}$ &  \\
89 Her          & 2.7 10$^{-2}$ & & & total neb.\ \\
89 Her          & 1.4 10$^{-2}$ & 1.5 & 2.3 10$^{16}$ & disk \\
IRAS 18123+0511 & 4.7 10$^{-2}$ & 0.6 & 3 10$^{16}$ & prob.\ exp.\ comp. \\
AC Her          & 8.4 10$^{-4}$ & 0.7 & 1.1 10$^{16}$ & \\
R Sct           & 5 10$^{-2}$ & &  & total neb.\ (complex prof.) \\
R Sct           &  $\sim$6.9 10$^{-3}$ & $\sim$1 & $\sim$1.5 10$^{16}$
& disk \\ 
IRAS 19125+0343 & 1.3 10$^{-2}$ & 1 & 2.3 10$^{16}$ & prob.\ exp.\ comp. \\
IRAS 19157-0247 & 1.4 10$^{-2}$ & 0.7 & 3 10$^{16}$ & \\
IRAS 20056+1834 & 10$^{-1}$ & & & total neb.\ (complex prof.) \\
IRAS 20056+1834 &  $\sim$2.5 10$^{-2}$ & $\sim$0.6 & $\sim$1.7
10$^{16}$ & disk \\ 
R Sge           & $<$ 9 10$^{-3}$ & $<$0.3 & $<$7 10$^{15}$ &  \\
\hline
 & & & & \\
IRAS 08544-4431 & $\sim$ 7.7 10$^{-3}$ & 2.2 & 1.8 10$^{16}$ & from 
\doce\ data \\
IW Car          & $\sim$ 5.3 10$^{-3}$ & 1.3 & 2 10$^{16}$ & from 
\doce\ data \\
HD 95767        & $\sim$ 1.2 10$^{-3}$ & 0.6 & 1.3 10$^{16}$ & from 
\doce\ data \\
HD 108015       & $\sim$ 2.3 10$^{-2}$ & 1.2 & 3 10$^{16}$ & from 
\doce\ data \\
\hline

\end{tabular}
\end{center}
\end{table}

The results of our mass estimate are shown in Table 5. We also give
comments on the estimate meaning and reliability.  Some sources deserve
a detailed discussion:

The rotating disk of the Red Rectangle has been well studied, 
see Sects.\ 1 and 4 and Bujarrabal et al.\ (2005, 2013). This object
has served in fact as a prototype to allow our study of other less
well-known nebulae. The mass derived in those studies, $M_{\rm mol}$
$\sim$ 6 10$^{-3}$ \ms, is completely compatible with that obtained
here (6.6 10$^{-3}$ \ms). It is important to note that the contribution
to the \trece\ low-$J$ profiles due to expanding gas components (which
is obvious in \doce\ lines, see Sect. 4) is probably minor. Bujarrabal
et al.\ (2013) derived, from their high-$J$ \doce\ profiles, a mass for
the outflowing gas in this object of about 10$^{-3}$ \ms, which could
be included in the slightly larger value of the total mass derived
here.

In 89 Her, the $\jdu$ maps show that the emission from velocities
separated by more than 3 -- 4 \kms\ from the line center comes only
from expanding gas. Accordingly, together with our standard estimate of
the mass, we have performed another one retrieving a degree 2 baseline
from the profiles that eliminates this spectral component. The result
is also given in Table 5.  Again our estimate is compatible with those
deduced by Bujarrabal et al.\ (2007), $\sim$ 10$^{-2}$ \ms.

The fact that the mass values derived here for the Red Rectangle and 89
Her are very similar to those deduced from completely independent (and
more detailed) estimates is a proof of the reliability of our simple
method.

IRAS\,18123+0511 and IRAS\,19125+0343 show profiles with a narrow core
and line wings. The wings are relatively intense in \doce\ \jdu, but
the profiles are much narrower in the optically thin \trece\ lines. It
seems therefore that in these sources there is a significant
contribution of an expanding component, but it is probably
  negligible in \trece\ \juc\ and so our standard mass
  calculation, only considering the contribution of the disk, is
  likely reliable for them.

R Sct and IRAS\,20056+1834 show composite profiles that suggest that
the contribution from a heavy outflow is important, in fact dominant,
in the resulting relatively wide profiles (10 -- 15 \kms). In both
cases, there is a narrow component, about 5 \kms\ wide, that could come
from a rotating disk. We have performed mass calculations also for this
disk component, trying to isolate in the observed profiles the
  contribution of the narrow features (in a similar way as for
  89 Her), but the estimate of the corresponding intensity is not
straightforward; values are given in Table 5. In any case, we think
that those values, 7 10$^{-3}$ \ms\ for R Sct and 2.5 10$^{-2}$
\ms\ for IRAS\,20056+1834, are the best estimates of the disk mass in
these nebulae. The total mass values deduced for these objects, close
to 0.1 \ms, are the highest in our sample, and seem dominated in fact
by the putative expanding component. We must keep in mind that these
mass values may be overestimated by about a factor 2 if the
characteristic rotational temperature in the dominant expanding
component is lower than that deduced here for rotating disks, as
happens for the (fast) outflowing components of more massive PPNe, in
which temperatures of about 20 K are often found (Bujarrabal et
al.\ 2001).

The rest of the detected sources show narrow profiles that strongly
suggest that the emission from an equatorial disk dominates and that we
are reasonably calculating the mass of this rotating component

In DY Ori, we do not detect the \trece\ \juc\ transition, deriving
$M_{\rm mol}$ $<$ 3 10$^{-3}$ \ms. But, from \trece\ \jdu\ (see
Sect.\ 5.2), we find $M_{\rm mol}$ \gsim\ 1.6 10$^{-3}$ \ms, so we can
safely assign a value of $M_{\rm mol}$ $\sim$ 2 10$^{-3}$ \ms\ to
this source.  For RV Tau and R Sge we derive just upper limits to the
molecular-gas mass, we find respectively for these sources $M_{\rm
  mol}$ $<$ 8 10$^{-3}$ \ms\ and $M_{\rm mol}$ $<$ 9 10$^{-3}$ \ms,
which are not much smaller than the results obtained in detected
sources.  From the \trece\ \juc\ limit found in U Mon, which is a
relatively nearby source, we deduce $M_{\rm mol}$ $<$ 9 10$^{-4}$ \ms.
This is one of the lowest mass values we have found in our sample; the
lowest actual measurement is that derived for HR\,4049, 6 10$^{-4}$
\ms.

These nebular mass values are significantly smaller than those found in
{\em standard} PPNe, often $\sim$ 0.2 -- 1 \ms\ (Sect.\ 1). This result
holds even if we take into account the uncertainty in the disk mass
determination, of about a factor 2 in general cases and about a factor
3 in sources with poorly known distance (Sect.\ 5.3.1).

We have also given mass estimates for the outflowing components, but
they are in general uncertain, as these outflows are not well studied
yet. In the Red Rectangle and 89 Her, however, relatively good
estimates of the outflowing gas mass can be obtained, from our
calculations and previous works (Bujarrabal et al.\ 2007, 2013). For
the Red Rectangle we conclude that the outflowing gas mass is
  about 
10$^{-3}$ \ms\ and the expansion velocity is about 10 \kms; for 89
Her we find a mass of about 10$^{-2}$ \ms\ and a typical velocity of 5
\kms. The linear momenta of the outflows in these sources is then of
about 2 10$^{36}$ g\,cm\,s$^{-1}$ and 10$^{37}$ g\,cm\,s$^{-1}$,
respectively, and their kinetic energy is $\sim$ 2 -- 5 10$^{42}$
erg. The linear momentum that can be delivered per unit time by
the stellar luminosity (in all directions and neglecting photon
trapping effects, i.e., $L/c$) is 2.5 10$^{34}$
g\,cm\,s$^{-1}$\,yr$^{-1}$ for the Red Rectangle and 4 10$^{34}$
g\,cm\,s$^{-1}$\,yr$^{-1}$ for 89 Her. (We recall that, as usual in
  this kind of comparisons, we are not exactly dealing with the linear
  momentum, but with a kind of {\it scalar linear momentum} integrated in
  all directions.) This shows that, contrary to
the case of most {\em standard} PPNe (Bujarrabal et al.\ 2001), the
linear momentum and energy carried by the bipolar outflows in these
objects is very moderate and stellar radiation could provide them in a
reasonable time (in comparison with the expected ejection times). Note
that the mass/radiation momentum ratio is distance independent.
We can reach a similar conclusion for the other objects that probably
show bipolar outflows (identified from the relatively intense line
wings, present, at least, in IRAS\,18123+0511, IRAS\,19125+0343, AI
CMi, and IRAS\,08544-4431), since the outflowing gas mass and velocity
in them, although less well known, cannot be significantly higher than
for the Red Rectangle and 89 Her. 

The momenta of the outflows of R Sct and IRAS\,20056+1834, which seem
the dominant nebular components, are around 5 -- 10 times higher than
for the Red Rectangle and 89 Her, but still far lower than those
typical of {\em standard} PPNe (Bujarrabal et al.\ 2001).  

R Sct and IRAS\,20056+1834 are in some way intermediate between the
low-luminosity post-AGB objects studied here and the (luminous and
massive) {\em standard} PPNe and young PNe; indeed the luminosity of R
Sct is as high as 10$^4$ \ls\ (that of IRAS\,20056+1834 is not well
known, Table 2). There are also PPNe with low luminosities ($\sim$
10$^3$ \ls) that do not show NIR excess nor disk-like CO profiles (but
significantly broader ones), like IRAS\,19500-1709 and
IRAS\,23304+6147, see Bujarrabal et al.\ (2001). The nebular masses in
these objects were found to be low, $\sim$ 10$^{-2}$ \ms, not
significantly larger than those of our objects (but note that these
nebulae are expanding at high velocity). 

Finally, we have estimated the disk mass in the southern sources we
have observed with APEX, also using a SEST observation from Maas et
al.\ (2003), see details in Appendix B. No \trece\ data are available
for those sources. The mass values are then derived after converting
those derived from \doce\ \jdu\ intensities to the values would derive
from \trece\ \juc\ data, using an empirical relation between results
from both lines for northern sources observed with the 30m telescope,
and taking into account the different telescope beam widths
(Sect.\ 3). In these calculations, we adopted a \doce\ abundance
  of 2 10$^{-4}$, but the assumed value has no effect on the finally
  derived mass, since we used an empirical relation to translate the
  obtained results to mass values equivalent to those obtained from
  \trece\ lines. Because of the high optical depth of \doce\ lines, the
  derivation of reliable values of the isotopic abundance ratio
  requires detailed modeling case by case, which is out of the scope of
  this paper; the low isotopic abundance ratio assumed here
  is compatible with previous studies of theses objects (Bujarrabal et
  al.\ 1990, 2013). These approximations obviously add uncertainty to
our estimates. The results are also given in Table 5, last four lines.
We didn't try to deduce upper limits for the undetected southern
objects, because of this uncertain conversion and the other
observational problems discussed in Sect.\ 3.2 and Appendix B.

\subsubsection{Uncertainties in the mass estimates}

The major sources of uncertainty in our estimates come from the
assumptions of $T_{\rm rot}$, less than a factor of two
(Sect.\ 5.2), and of $X$(\trece). We think that the uncertainty in our
assumption on $X$(\trece) is also smaller than a factor of two,
since existing estimates yield values of this parameter quite similar
to our adopted one. In AGB stars, $X$(\doce) is found to vary between
about 2--4 10$^{-4}$ in O-rich stars and 6--8 10$^{-4}$ in carbon-rich
stars (see e.g.\, Teyssier et al.\ 2006, Ramstedt et al.\ 2008, and
references therein), but the \doce/\trece\ abundance ratio is higher in
carbon-rich stars, resulting in a similar value of the
\trece\ abundance, which is usually found to be $X$(\trece) $\sim$ 2
10$^{-5}$ (e.g.\ Kahane et al.\ 1992, 2000, Bujarrabal et al.\ 1994,
Sch\"oier \& Olofsson 2000, Sch\"oier et al.\ 2011, etc).  Studies of
molecular gas in PNe and PPNe, often give values of $X$(\doce) slightly
lower than for AGB stars, but values $X$(\trece) $\sim$ 2 10$^{-5}$ are
usually compatible with the observations (e.g.\ Bujarrabal et
al.\ 2001, 2013, Woods et al.\ 2005, Soria-Ruiz et al.\ 2013,
etc). $X$(\trece) $\sim$ 2 10$^{-5}$ is in particular the value deduced
for the Red Rectangle by Bujarrabal et al.\ (2013) and used in their
systematic study of PPNe by Bujarrabal et al.\ (2001).  In the case
that we assume a different value for $X$(\trece), the nebular mass we
deduce here is easily scaled following $M_{\rm mol}$ $\propto$
1/$X$(\trece).  Both uncertainties, due to the excitation conditions
and due to the assumed abundance, are not expected to be correlated, so
we can assume a total uncertainty of about a factor 2. We must keep in
mind that our estimates of the nebular mass concern only the
molecule-rich component. The central stars of our sources (Tables 1, 2)
are often not hot enough to significantly photodissociate CO, but the
presence of a hotter companion or the interstellar UV field could yield
photodissociation of molecules (respectively in inner and outer disk
regions). Therefore, our estimates are always a lower limit to the
total nebular mass, which may include components of (mostly) atomic gas
with significant amounts of mass, in particular extended haloes, not
probed by our observations.

Another, sometimes difficult to quantify source of uncertainty is the
distance estimate. In the worst cases, we have deduced a distance value
just assuming a typical luminosity of 
3000 \ls\ (Sect. 2.1). Since the luminosity of post-AGB stars is not
expected to differ by more than a factor 2--3 from that assumed value,
we expect in those cases errors in the distance by factors $\sim$ 1.5
and an error in the mass not larger than a factor 2--3. Taking into
account the sources of uncertainty mentioned above, we derive a total
uncertainty in the mass determination of a factor 3--4 for objects in
which the distance is poorly known.

\subsection{Rough estimates of the typical sizes of the disks}

Following a reasoning similar to those developed in previous
subsections, it is possible to estimate the typical sizes of the disks
responsible for the detected lines. We will assume that the low-$J$
lines are thermalized and that the \doce\ \jdu\ line is opaque. The
first assumption very probably holds in view of the low Einstein
coefficients of these low transitions, \lsim\ 10$^{-6}$ s$^{-1}$. The
\doce\ \jdu\ line is very probably opaque, since the measured ratios
between this line and the more optically thin \doce\ \juc\ and
\trece\ lines is much lower than expected for optically thin
emissions. Under these assumptions, the main-beam brightness
temperatures at the line center must be aproximately equal to the ratio
between the solid angles occupied by the source and the telescope beam,
multiplied by the typical temperature (here assumed to be 50 K). This
estimate is rough because very probably the temperature and opacity
vary across the source and, therefore, we are deriving an {\em average}
or {\em characteristic } size, disregarding for instance the disk
inclination. But we think that it will give an idea of a parameter that
is unknown for all sources, excepted the Red Rectangle and 89 Her. Once
we assume a value of the distance (Sect.\ 2, Table 1), we can also
derive a typical linear size. Values derived in this way are summarized
in Table 5. In the sources in which two components are considered,
  we only derive the size of the disk, that of the expanding component
  being still more uncertain. In any case, we must keep in mind that
  the expanding component could always contribute to the size derived
  in this way.

Typical derived sizes range between about 0\farcs5, for weak sources
with $T_{\rm mb}$ $\sim$ 0.1 K, and 2$''$ for intense sources with
$T_{\rm mb}$ $\sim$ 1.5 K. 

\section{Conclusions}

We have studied a sample of 24 post-AGB stars that show NIR excess
indicative of the presence of hot dust, which was proposed to be placed
in an inner rotating disk. These stars are systematically binaries and
often show lower luminosities and temperatures than those of the {\em
  standard} well-studied protoplanetary and young planetary nebulae,
see Sects.\ 1 and 2. 

Fourteen sources were observed with the IRAM 30m telescope in the
\doce\ and \trece\ \jdu\ and \juc\ transitions. Eleven of these objects
(the Red Rectangle, 89 Her, DY Ori, HR\,4049, IRAS\,18123+0511, AC Her,
IRAS\,19125+0343, IRAS\,19157-0247, AI CMi, R Sct,and IRAS\,20056+1834)
were clearly detected.
See data in Table 3 and Figs.\ 1 to 11 and A.1 to A.3.

Ten southern sources were observed in the \doce\ \jtd\ and
\jdu\ transitions with APEX, see Sect.\ 2.2, Appendix B, Table 4 and
Figs.\ B.1 to B.10. Four sources (IRAS\,08544-4431, IW Car, HD\,95767,
and HD\,108015) are detected and show line profiles similar to those of
the northern sources.  Because of the poorer S/N ratios attained in
this southern sample and the presence of strong interstellar
contamination in several sources, the detection rate is in this case
less meaningful.

Most of the observed line profiles are very similar to those of the Red
Rectangle, the only post-AGB source in which the presence of
rotation in the molecule-rich nebula is well proven (Bujarrabal et
al.\ 2005). The single-dish profiles of the Red Rectangle (Fig.\ 2)
show characteristic line profiles composed of a central narrow peak
plus low-velocity wings, similar to those expected from rotating disks
and found in disks orbiting young stars, and very different of the wide
profiles found in AGB stars and other PPNe (Sect.\ 4). It is known that
the Red Rectangle and 89 Her also show a gas component in axial
expansion, occupying a bipolar structure, that is responsible (at least
partially) for the relatively intense emission at LSR velocities
farther than about $\pm$ 5 \kms\ from the systemic one.  In view of the
disk-like line profiles, very similar to those of the Red Rectangle, we
conclude that the dominant component in most of our nebulae is a
rotating equatorial disk (Sect.\ 4).

AI CMi, IRAS\,18123+0511, and IRAS\,19125+0343 show disk-like profiles
with relatively intense wings that strongly suggest that gas in
expansion is also present in them. The presence of an expanding
component is also probable in the southern source IRAS\,08544-4431. In
R Sct and IRAS\,20056+1834 the profiles are more complex, but showing a
narrow component, which suggest that a rotating disk is present but the
total emission is dominated by that of expanding components.

We stress the high detection rate of rotating disks among the objects
well studied with the 30m telescope. In fact, the limits found for the
undetected sources (RV Tau, U Mon, and R Sge) do not lead to upper
limits of the mass and size of the disks (see below) much lower than
the values found in the detected nebulae, strongly suggesting that
nondetection in these cases is just due to the attained sensitivity.

We have calculated the mass of the CO-emitting gas, assuming the line
emission properties expected in rotating disks, mostly deduced from
previous studies of the Red Rectangle and 89 Her, see Sect.\ 5. We
argue that the expected uncertainty in the mass determination probably
do not exceed a factor 2 (a factor 3-4 when the distance is very poorly
known). The derivation of the mass is performed from data of
\trece\ \juc\ when it was detected, because this line is less affected
by opacity effects and the relatively large telescope beam at this
frequency prevents nondetection of extended components. For our
southern sources observed with APEX, in which data of this line are not
available, we derived mass estimates assuming an empirical relation we
found between results obtained from \trece\ \juc\ and
\doce\ \jdu\ (Appendix B); we must keep in mind that these results are
significantly more uncertain than for the other objects. A summary of
the mass values is shown in Table 5. We tried to separate the
contributions of the expanding and rotating components in some objects:
the Red Rectangle and 89 Her, for which we have information on the
contribution of both components from other works, and R Sct and
IRAS\,20056+1834, in which the putative expanding component seems
dominant.

The mass of the rotating disks derived here typically ranges between
10$^{-3}$ and 10$^{-2}$ \ms. Higher values are only found in objects in
which a significant contribution of an expanding component is probable;
for R Sct and IRAS\,20056+1834 we deduce a total mass of 5 10$^{-2}$ --
10$^{-1}$ \ms, but we note that our procedure may overestimate the mass
values by up to a factor two in these cases (since we are assuming
emission properties typical of disks, see Sect.\ 5). Only HR\,4049, AC
Her, and U Mon show disk mass values somewhat smaller than 10$^{-3}$
\ms. U Mon was not detected and we derive in fact an upper limit of 9
10$^{-4}$ \ms. The mass upper limits derived for RV Tau and R Sge, our
two other significant nondetections, are higher than 10$^{-3}$ \ms.

The total and disk mass values derived in our objects are significantly
smaller than that derived for {\em standard}, well-studied PPNe and PNe
(CRL\,2688, CRL\,618, NGC\,7027, etc; Sect.\ 1), which show nebular
masses of about 0.2 -- 1 \ms\ (e.g.\ Bujarrabal et al.\ 2001). This
result holds in spite of the uncertainties in the mass estimates. The
velocities found in our objects, even those of the expanding components
and taking into account projection effects, are always very moderate,
\lsim\ 10 \kms, significantly smaller than those usual in the above
mentioned {\em standard} nebulae, which can be larger than 100 \kms.

Although the existing information on the expanding components is still
poor, we derive low mass values for the best studied cases (the Red
Rectangle and 89 Her, Sect.\ 5.3), between 10$^{-3}$ and a 10$^{-2}$
\ms. The typical expansion velocities are found to range between 5 and
10 \kms. Others objects studied here, notably AI CMi, IRAS\,18123+0511,
IRAS\,19125+0343, and IRAS\,08544-4431, probably present similar values
of these parameters. Consequently, the linear momenta of the outflows
in these objects are low, comparable to what stellar radiation pressure
could deliver in reasonable times of a few hundred years and much lower
than those of the very massive and fast outflows of {\em standard} PPNe
(Bujarrabal et al.\ 2001). Even for R Sct and IRAS\,20056+1834, in
which the outflowing component is dominant, we derive moderate linear
momentum values (Sect.\ 5.3). We stress that this result does not show
that radiation pressure is actually powering these bipolar outflows; in
fact, we recall that the studied objects present the main properties
(binarity, rotating disk, possible presence of reaccretion) that are
required to explain bipolar nebulae and flows by interaction between
previously ejected shells and post-AGB jets, see Sect.\ 1. It is
remarkable that there is no clear evidence, to our knowledge, of disks
in rotation in PPNe with high luminosity and mass, although, precisely
in these objects, the acceleration of their powerful outflows seems
to require reaccretion from a disk.

It is possible to derive estimates of the disk extent from the
\doce\ \jdu\ main-beam brightness temperatures measured in our objects,
by assuming that this line is optically thick and thermalized, which
are acceptable assumptions (Sect.\ 5, Table 5). Typical derived sizes
range between about 0\farcs5 and 2$''$, equivalent to typical linear
sizes between 5 10$^{15}$ and 3 10$^{16}$ cm.
These CO-rich disks are therefore significantly larger than the very
compact disks ($<$ 10$^{15}$ cm) supposed to be responsible for the NIR
excess of these objects, but they are much smaller than {\em standard}
PPNe and young PNe, which show total sizes larger than 10$^{17}$ cm
(e.g.\ Bujarrabal et al.\ 1988, 2001).

In summary, the high detection rate in our observations of extended
rotating disks indicates that such structures are common in post-AGB
stars, being present in all or almost all post-AGB sources showing a
significant NIR excess (which are a good fraction of catalogued
post-AGB stars, Sect.\ 1). Our results confirm that the compact dust
components responsible for this NIR excess are in rotation and very
probably coincide with the innermost regions of the CO emitting
disks. We derive disk (and total) mass values, which range between
10$^{-3}$ and 10$^{-2}$ \ms, significantly smaller than the mass
of well-known {\em standard} PPNe.  Two of our objects, the Red
Rectangle and 89 Her, are known to present also low-velocity bipolar
outflows; in others, our single-dish profiles strongly suggest the
presence of similar components. In general, the outflows are a minor
component of the total nebula, except probably in objects with
complex profiles. 

The main properties of the different kinds of post-AGB objects, notably
the binarity and the nebular mass and dynamics, as well as the possible
formation of planetary nebulae from them, will be the subject of a
forthcoming work.
 

\begin{acknowledgements}
This work has been supported by the Spanish MICINN, program CONSOLIDER
INGENIO 2010, grant ``ASTROMOL" (CSD2009-00038). We are grateful to the
anonymous referee of the paper for his/her helpful comments. This work
has made extensive use of the SIMBAD database
(http://simbad.u-strasbg.fr/simbad/sim-fid).
\end{acknowledgements}

{}

\newpage
\appendix

\section{Additional 30m-telescope CO data}

In this Appendix we show our 30m telescope data of RV Tau, U Mon, and R
Sge, the only objects of our northern sample which were not
detected. Wide spectral bands are shown, centered on the stellar
velocity expected from optical observations. See a summary of the
observational parameters in Sect.\ 2 and Table 3.

   \begin{figure}
   \centering \rotatebox{0}{\resizebox{8.5cm}{!}{ 
\includegraphics{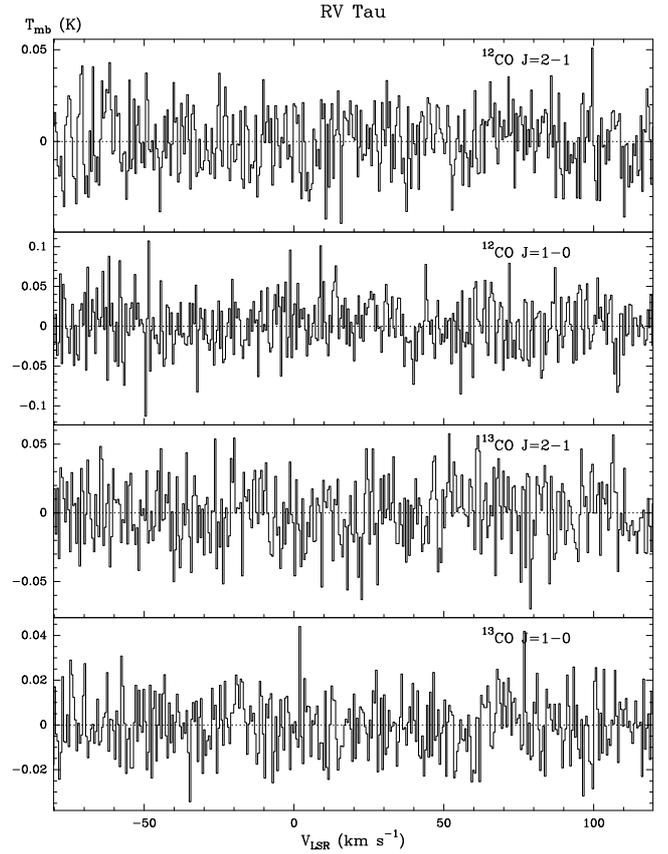}
}}
   \caption{30m telescope observations of RV Tau. No line is detected.}
    \end{figure}

   \begin{figure}
   \centering \rotatebox{0}{\resizebox{8.5cm}{!}{ 
\includegraphics{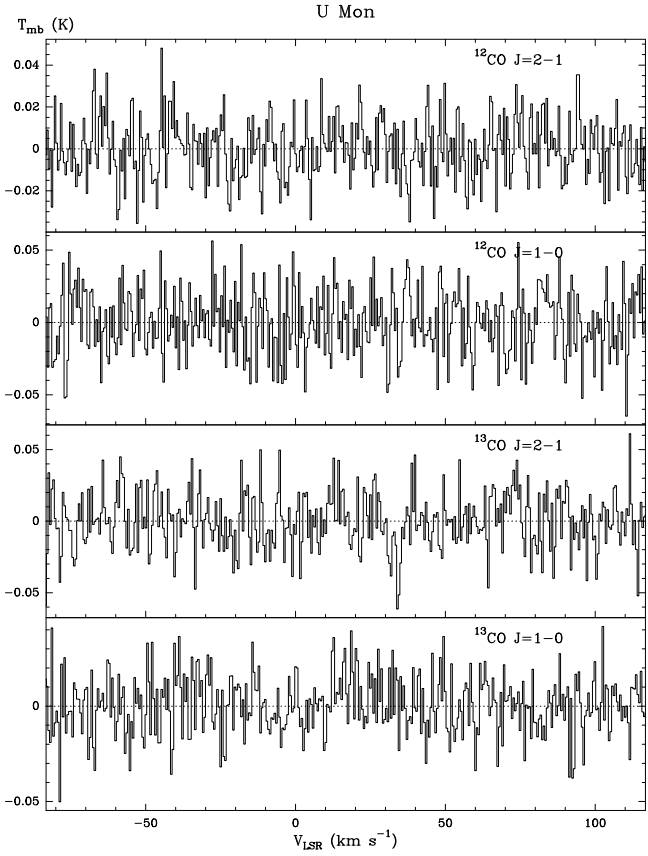}
}}
   \caption{30m telescope observations of U Mon. No line is detected.}
              \label{}%
    \end{figure}

   \begin{figure}
   \centering \rotatebox{0}{\resizebox{8.5cm}{!}{ 
\includegraphics{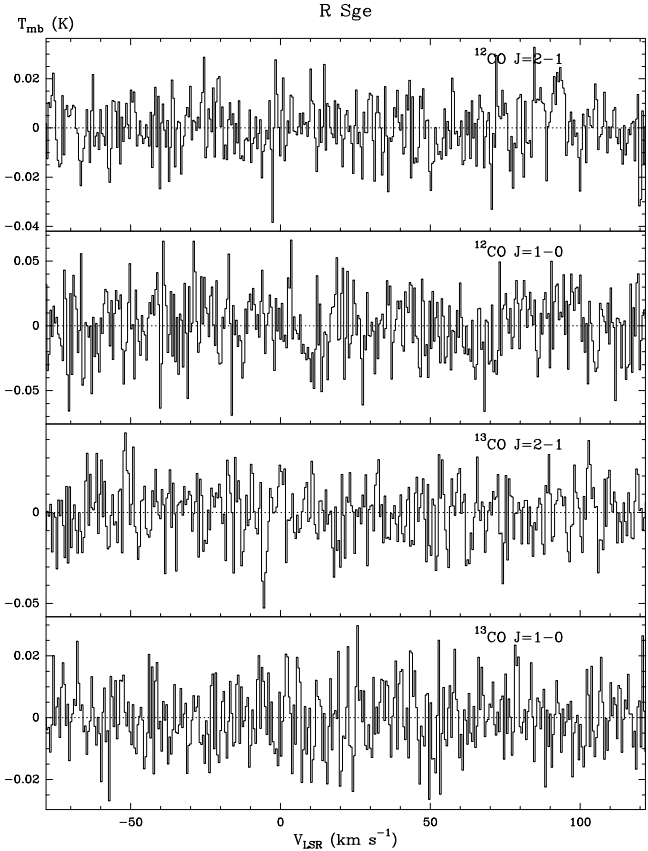}
}}
   \caption{30m telescope observations of R Sge. No line is detected.}
    \end{figure}

\newpage

\section{APEX observations of southern sources: spectra and disk mass
  estimates}

We have also observed a sample of southern post-AGB stars using APEX in
the \doce\ \jtd\ and \jdu\ transitions, see the source sample in Table
2 (we give the same parameters as in Table 1 for the northern
sources). Because of the lack of \trece\ observations and the poorer
quality of the data, which include in particular spectra dominated by
interstellar (IS) contamination, these results are not analyzed with
the same detail as those of northern sources. When strong IS
contamination is present it is even impossible trying to give a
reasonable upper limit to the post-AGB source emission, since the IS
contribution can be very intense, both positive and negative,
and practically impossible to predict.  The observational procedure is
described in Sect.\ 3.2, a summary of the line parameters is given in
Table 4 (similar to that given in Table 3), and the observed profiles
are given in Figs.\ B.1 to B.10.

In order to derive mass values for the disk in the detected sources, we
have first compared the mass values derived from the
\trece\ \juc\ integrated main-beam temperatures (our best estimates, in
Table 5) with those derived \doce\ \jdu\ (very probably
underestimates), following in both cases the method described in
Sect.\ 5. We have not considered the mass derived for R Sct and
IRAS\,20056+1834, in which the contribution of an additional component,
probably in expansion, is largely dominant. The result is shown in
Fig.\ B11. As we see, the mass values derived from \doce\ \jdu\ are
approximately ten times smaller than those obtained from \trece\ \juc,
which are the best estimates. We will accordingly estimate the mass of
the southern sources from their \doce\ \jdu\ intensities and correct
them by a factor 10. In the calculation of the disk mass values from
the SEST and APEX data we take into account that the APEX beam solid
angle (at 230 GHz) is about 4.7 times larger than that of the 30m
telescope and that the beam of the SEST is about 4 times larger. The
(uncertain) mass estimates derived in this way for our four detected
southern sources are given in Table 5 (last four lines).

   \begin{figure}
   \centering \rotatebox{0}{\resizebox{8.5cm}{!}{ 
\includegraphics{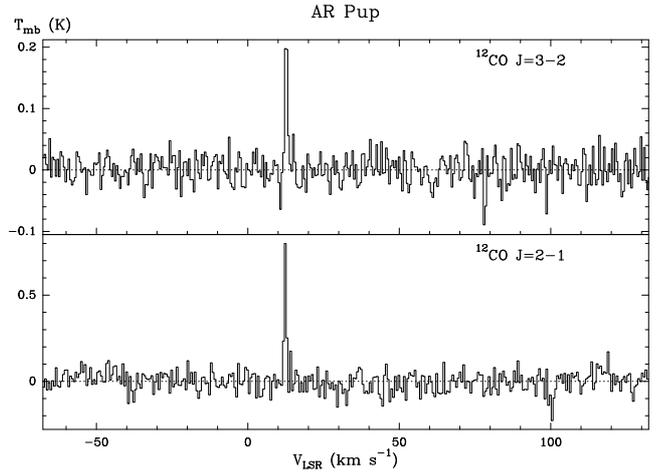}
}}
   \caption{APEX observations of AR Pup. No circumstellar line is
     detected, only a narrow interstellar feature.}
              \label{}%
    \end{figure}

   \begin{figure}
   \centering \rotatebox{0}{\resizebox{8.5cm}{!}{ 
\includegraphics{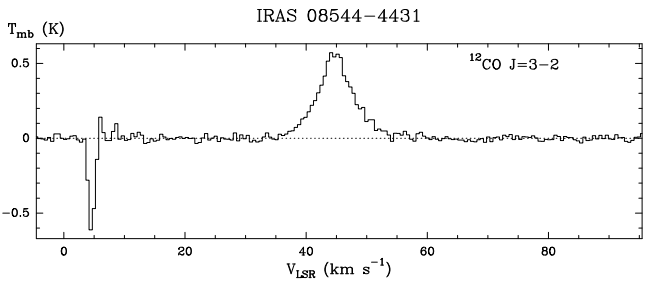}
}}
   \caption{APEX observations of IRAS\,08544-4431. \doce\ \jtd\ is
     detected (see previous data by Mass et al.\ 2003).}
              \label{}%
    \end{figure}

   \begin{figure}
   \centering \rotatebox{0}{\resizebox{8.5cm}{!}{ 
\includegraphics{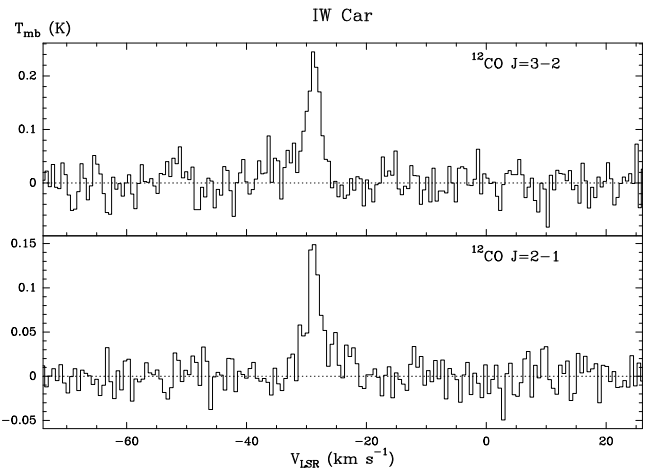}
}}
   \caption{APEX observations of IW Car. Both lines are detected.}
              \label{}%
    \end{figure}

   \begin{figure}
   \centering \rotatebox{0}{\resizebox{8.5cm}{!}{ 
\includegraphics{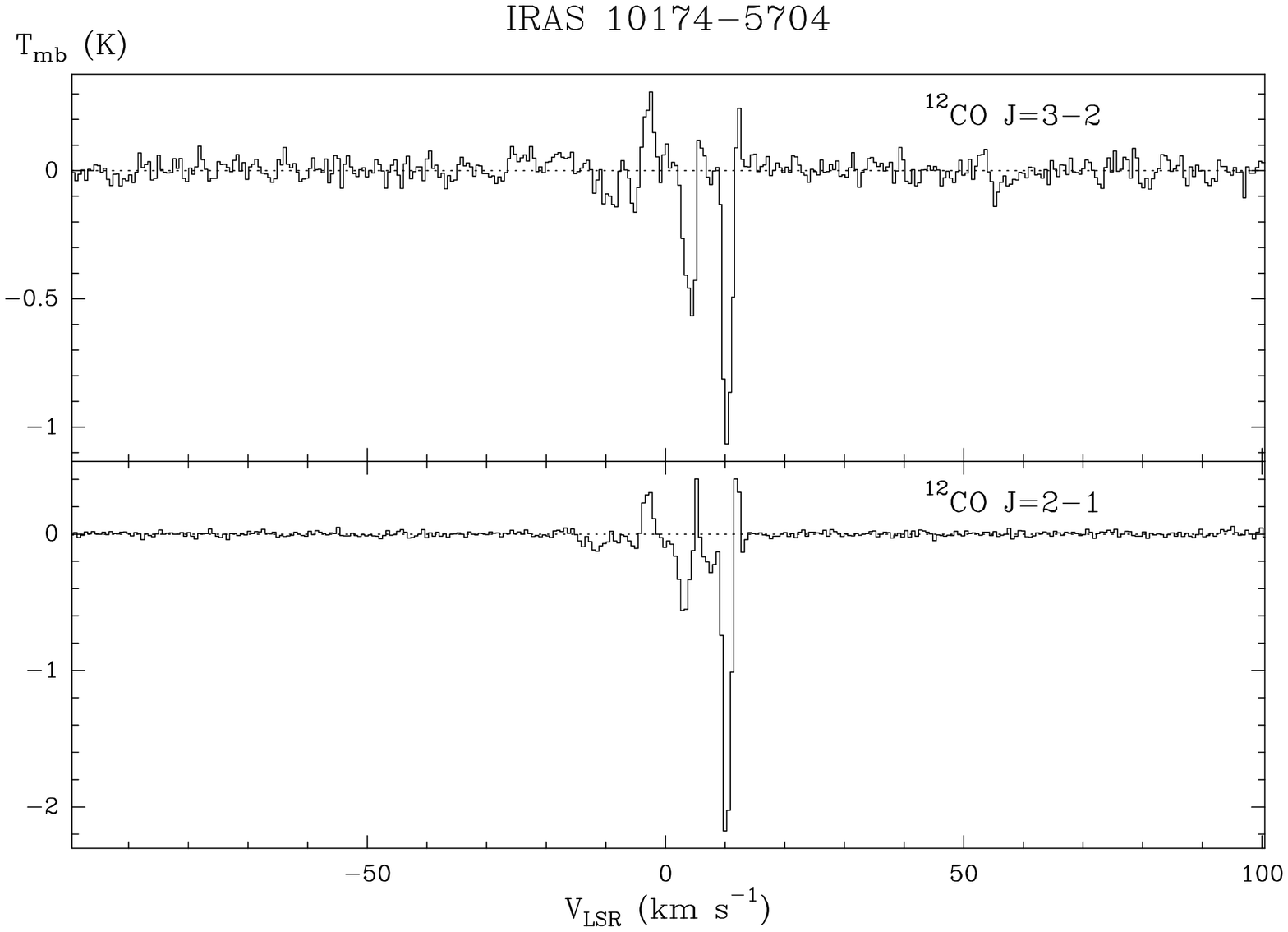}
}}
   \caption{APEX observations of IRAS\,10174-5704. Strong interstellar
     contamination at the relevant velocities prevents any conclusion on
     the emission from our source.}
              \label{}%
    \end{figure}

   \begin{figure}
   \centering \rotatebox{0}{\resizebox{8.5cm}{!}{ 
\includegraphics{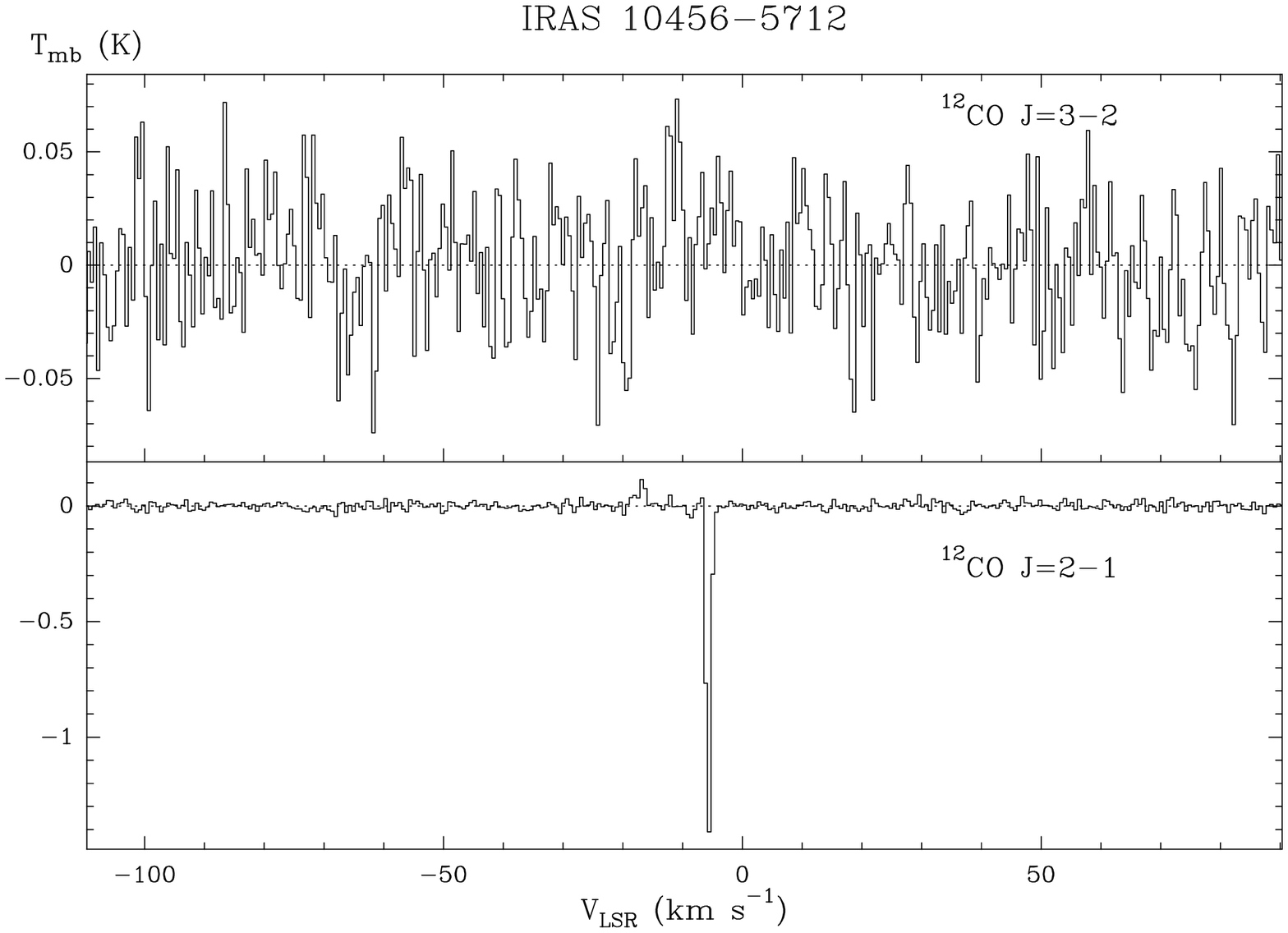}
}}
   \caption{APEX observations of IRAS\,10456-5712. Strong interstellar
     contamination at the relevant velocities prevents any conclusion on
     the emission from our source.}
              \label{}%
    \end{figure}

   \begin{figure}
   \centering \rotatebox{0}{\resizebox{8.5cm}{!}{ 
\includegraphics{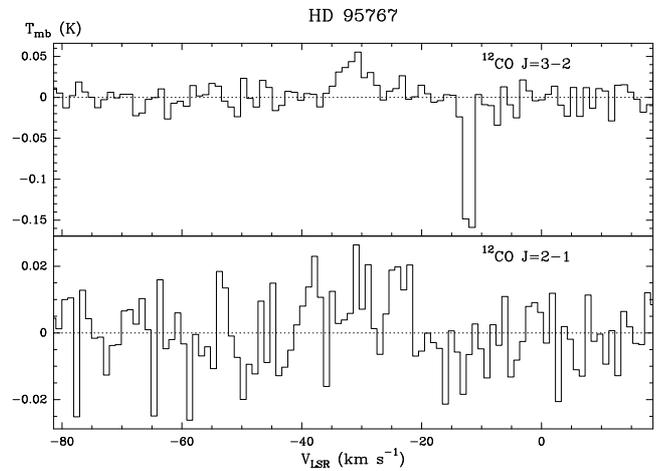}
}}
   \caption{APEX observations of HD\,95767. Emission is detected in
     \jtd, at the velocity found from optical spectroscopy, and
     tentatively in \jdu.}
              \label{}%
    \end{figure}

   \begin{figure}
   \centering \rotatebox{0}{\resizebox{8.5cm}{!}{ 
\includegraphics{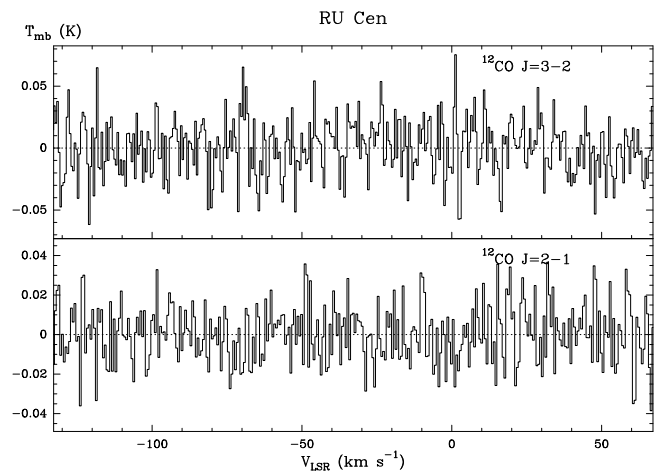}
}}
   \caption{APEX observations of RU Cen. No circumstellar line is
     detected.}
              \label{}%
    \end{figure}

   \begin{figure}
   \centering \rotatebox{0}{\resizebox{8.5cm}{!}{ 
\includegraphics{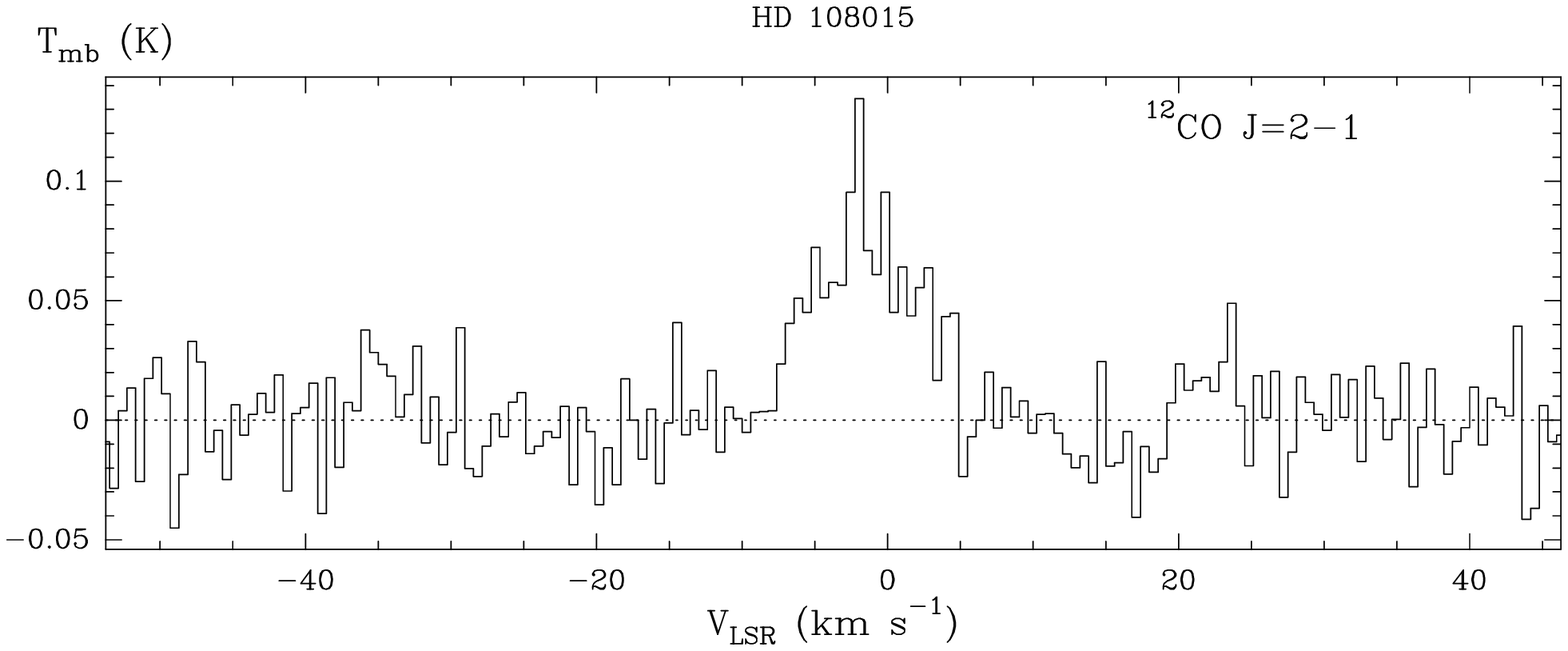}
}}
   \caption{APEX observations of HD\,108015. \doce\ \jdu\ is
     detected.}
              \label{}%
    \end{figure}

   \begin{figure}
   \centering \rotatebox{0}{\resizebox{8.5cm}{!}{ 
\includegraphics{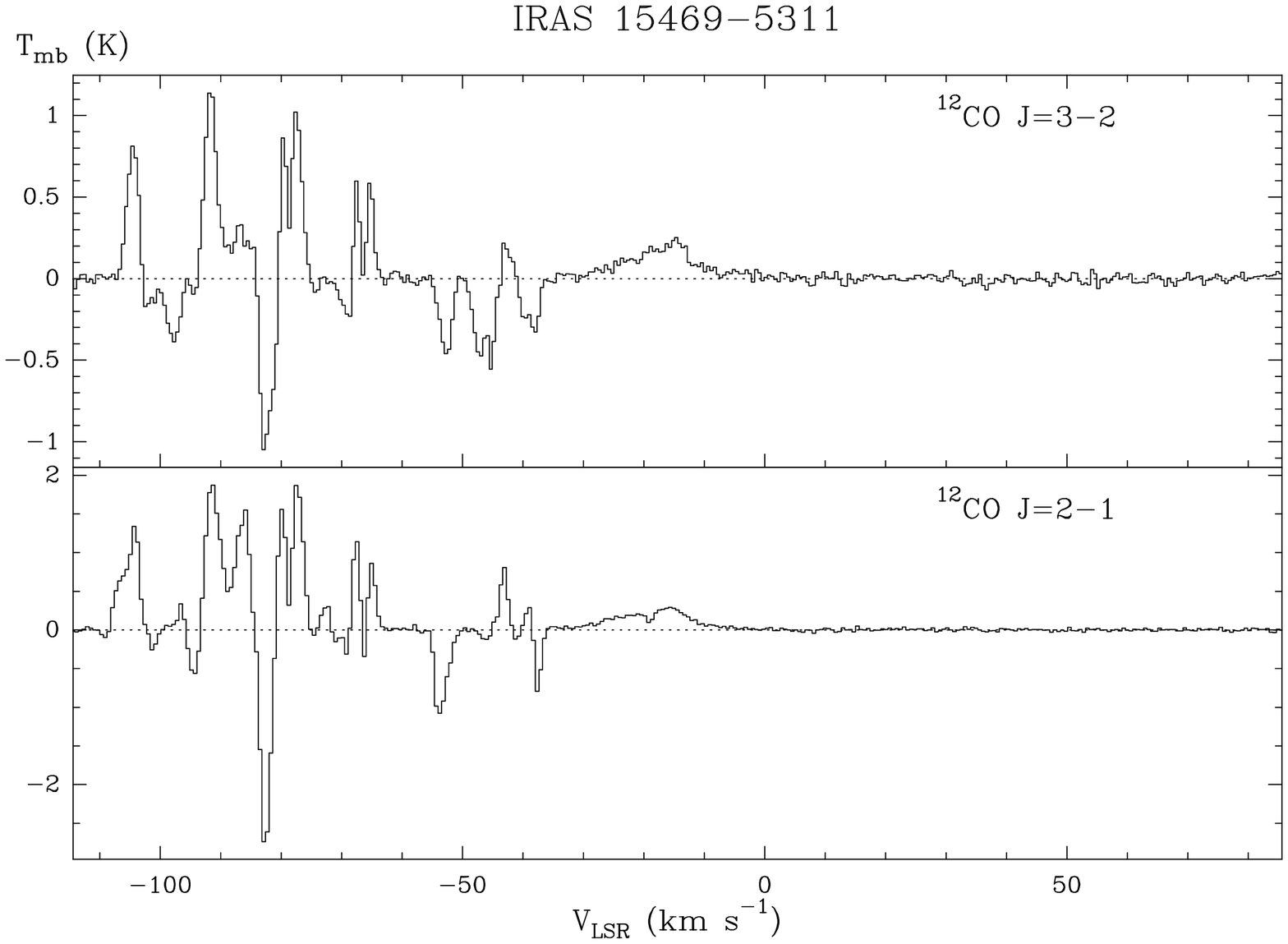}
}}
   \caption{APEX observations of IRAS\,15469-5311. Strong interstellar
     contamination at the relevant velocities prevents any conclusion on
     the emission from our source.}
              \label{}%
    \end{figure}

   \begin{figure}
   \centering \rotatebox{0}{\resizebox{8.5cm}{!}{ 
\includegraphics{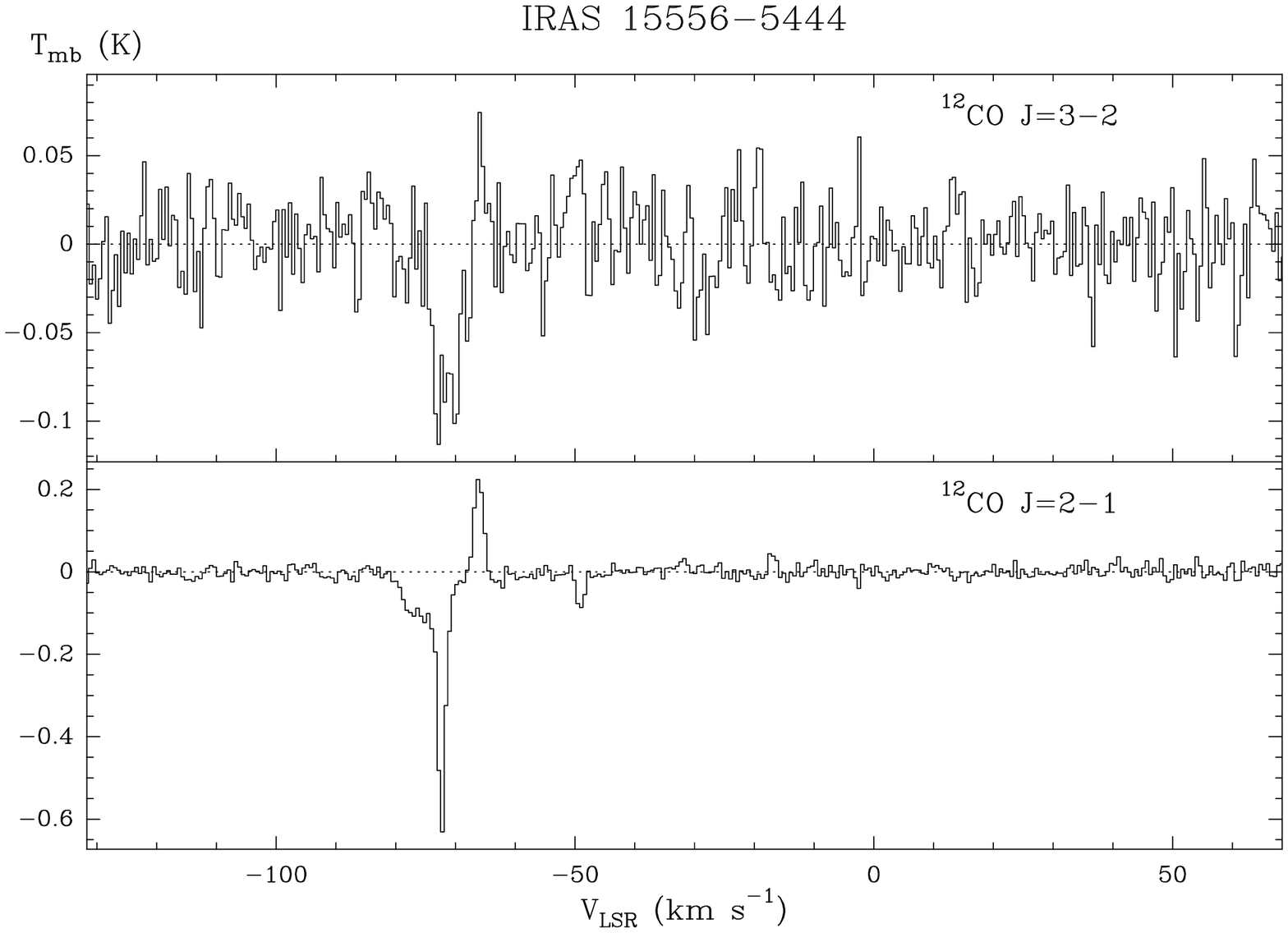}
}}
   \caption{APEX observations of IRAS\,15556-5444. Strong interstellar
     contamination at the relevant velocities prevents any conclusion on
     the emission from our source.}
              \label{}%
    \end{figure}

   \begin{figure}
   \centering \rotatebox{0}{\resizebox{8.5cm}{!}{ 
\includegraphics{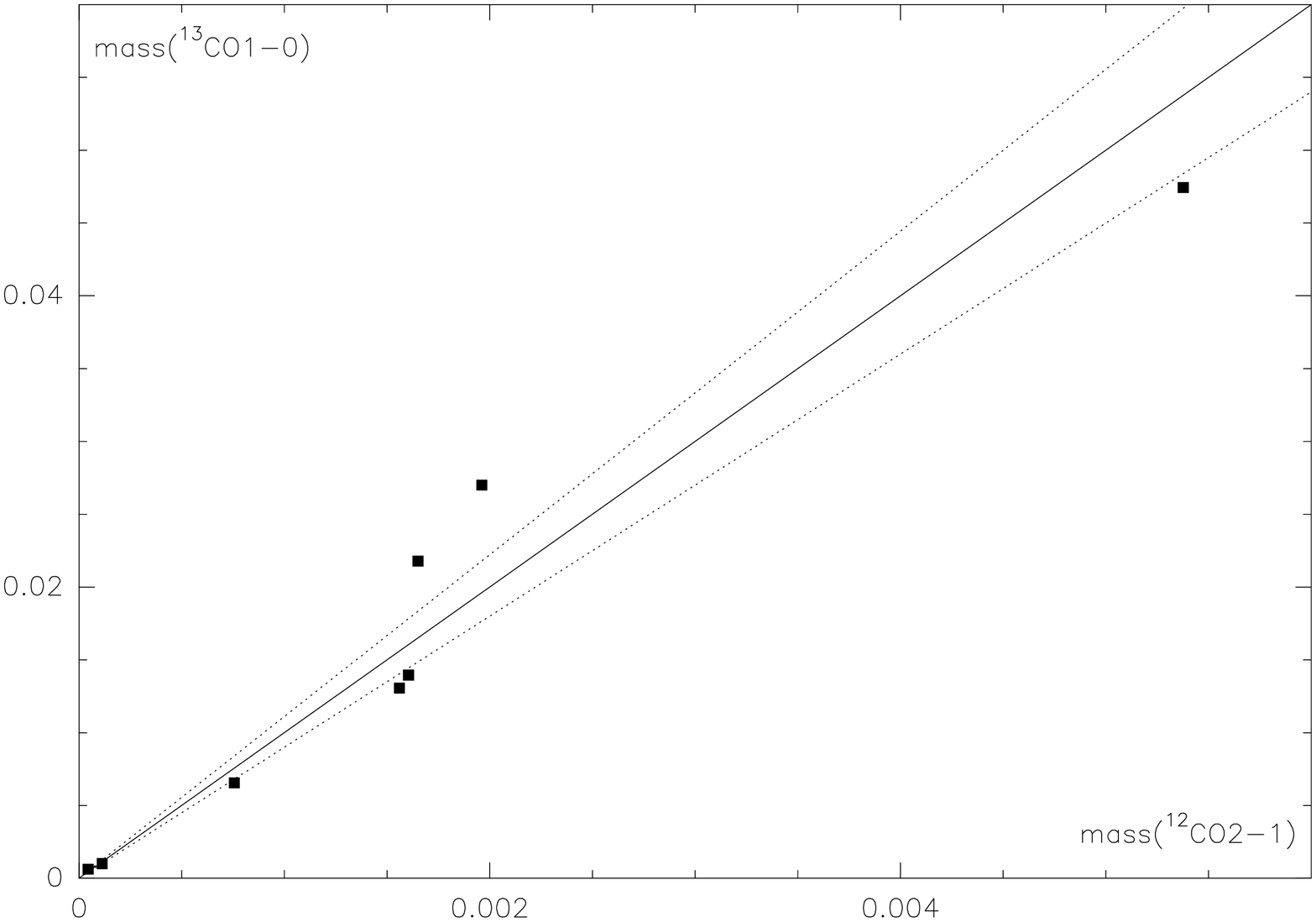}
}}
   \caption{Relation between the mass values (\ms) derived respectively from
     \trece\ \juc\ and \doce\ \jdu\ 30m telescope data, for the objects
     in which both determinations can be performed. Results from
     \trece\ \juc\ are expected to be the most accurate ones, those
     from \doce\ \jdu\ being probably underestimates. The continuous
     line represents a factor of ten between both estimates, and the
     other lines represent deviations of a 10\% from this ratio.}
\end{figure}


\begin{thebibliography}{}

\bibitem[Acke et al.(2013)]{2013A&A...551A..76A} Acke, B., Degroote,
  P., Lombaert, R., et al.\ 2013, \aap, 551, A76

\bibitem[Acker et al.(2012)]{2012RMxAA..48..223A} Acker, A., Boffin,
  H.~M.~J., Outters, N., et al.\ 2012, \rmxaa, 48, 223

\bibitem[Alcock et al.(1998)]{1998AJ....115.1921A} Alcock, C., Allsman, 
R.~A., Alves, D.~R., et al.\ 1998, \aj, 115, 1921 

\bibitem[Alcolea et al.(2007)]{alcolea07} Alcolea, J., Neri, R., \&
  Bujarrabal, V.\ 2007, \aap, 468, L41

\bibitem[Balick \& Frank(2002)]{2002ARA&A..40..439B} Balick, B., \&
  Frank, A.\ 2002, \araa, 40, 439

\bibitem[Bujarrabal et al.(1988)]{1988A&A...206L..17B} Bujarrabal, V.,
  Bachiller, R., Alcolea, J., \& Martin-Pintado, J.\ 1988, \aap, 206,
  L17
 
\bibitem[Bujarrabal et al.(1990)]{1990A&A...234..355B} Bujarrabal, V.,
  Alcolea, J., \& Bachiller, R.\ 1990, \aap, 234, 355

\bibitem[Bujarrabal et al.(2001)]{bujetal01} Bujarrabal, V.,
  Castro-Carrizo, A., Alcolea, J., \& S{\'a}nchez Contreras, C.\ 2001,
  \aap, 377, 868

\bibitem[Bujarrabal et al.(2005)]{bujetal05} Bujarrabal, V.,
  Castro-Carrizo, A., Alcolea, J., \& Neri, R.\ 2005, \aap, 441, 1031

\bibitem[Bujarrabal et al.(2007)]{bujetal07} Bujarrabal, V., van
  Winckel, H., Neri, R., et al.\ 2007, \aap, 468, L45

\bibitem[Castro-Carrizo et al.(2001)]{2001A&A...367..674C}
  Castro-Carrizo, A., Bujarrabal, V., Fong, D., et al.\ 2001, \aap,
  367, 674

\bibitem[de Ruyter et al.(2005)]{2005A&A...435..161D} de Ruyter, S.,
  van Winckel, H., Dominik, C., Watqers, L.~B.~F.~M., \& Dejonghe,
  H.\ 2005, \aap, 435, 161

\bibitem[Deroo et al.(2006)]{2006A&A...450..181D} Deroo, P., van
  Winckel, H., Min, M., et al.\ 2006, \aap, 450, 181

\bibitem[Fong et al.(2001)]{2001A&A...367..652F} Fong, D., Meixner, M.,
  Castro-Carrizo, A., et al.\ 2001, \aap, 367, 652

\bibitem[Frank \& Blackman(2004)]{2004ApJ...614..737F} Frank, A., \&
  Blackman, E.~G.\ 2004, \apj, 614, 737

\bibitem[Gielen et al.(2009)]{2009A&A...508.1391G} Gielen, C., van
  Winckel, H., Reyniers, M., et al.\ 2009, \aap, 508, 1391

\bibitem[Gielen et al.(2011)]{gielen11} Gielen, C., Bouwman, J., van
  Winckel, H., et al.\ 2011, A\&A, 533, 99

\bibitem[Huggins et al.(1996)]{1996A&A...315..284H} Huggins, P.~J.,
  Bachiller, R., Cox, P., \& Forveille, T.\ 1996, \aap, 315, 284

\bibitem[Guilloteau \& Dutrey(1998)]{1998A&A...339..467G} Guilloteau,
  S., \& Dutrey, A.\ 1998, \aap, 339, 467

\bibitem[Guilloteau et al.(2013)]{2013A&A...549A..92G} Guilloteau, S.,
  Di Folco, E., Dutrey, A., et al.\ 2013, \aap, 549, A92

\bibitem[G{\"u}sten et al.(2006)]{gusten2006} G{\"u}sten, R., Nyman,
  L.~{\AA}., Schilke, P., et al.\ 2006, \aap, 454, L13

\bibitem[Kahane et 
al.(1992)]{kahane92} Kahane, C. Cernicharo, J., G\'omez-Gonz\'alez, J.,
  Gu\'elin, M., 1992, \aap, 256, 235

\bibitem[Kahane et 
al.(2000)]{kahane00} Kahane, C. Dufour, E. Busso, M, et al., 2000,
  \aap, 357, 669

\bibitem[Kimura et al.(2012)]{2012A&A...541A.112K} Kimura, R.~K.,
  Gruenwald, R., \& Aleman, I.\ 2012, \aap, 541, A112

\bibitem[Klein et 
al.(2012)]{klein2012} Klein, B., Hochg{\"u}rtel, S., Kr{\"a}mer, I., et al.\ 2012, \aap, 542, L3 

\bibitem[Maas et al.(2003)]{2003A&A...405..271M} Maas, T., Van Winckel,
  H., Lloyd Evans, T., et al.\ 2003, \aap, 405, 271

\bibitem[Men'shchikov et al.(2002)]{men02}
Men'shchikov, A.B., Schertl, D., Tuthill, P.G., et al.\ 2002, A\&A, 393,
867

\bibitem[Pottasch(1984)]{1984ASSL..107.....P} Pottasch, S.~R.\ 1984, 
Astrophysics and Space Science Library, 107,  

\bibitem[Ramstedt et al.(2008)]{ramstedt08} Ramstedt, S., Sch\"oier,
  F., Olofsson, H., Lundgren, A.A., 2008, \aap, 486, 645

\bibitem[Kameswara Rao et al.(2002)]{2002MNRAS.334..129K} Kameswara Rao, 
N., Goswami, A., \& Lambert, D.~L.\ 2002, \mnras, 334, 129 

\bibitem[Risacher et 
al.(2006)]{risacher2006} Risacher, C., Vassilev, V., Monje, R., et al.\ 2006, \aap, 454, L17 

\bibitem[Sch{\"o}ier 
\& Olofsson(2000)]{2000A&A...359..586S} Sch{\"o}ier, F.~L., \&
  Olofsson, H.\ 2000, \aap, 359, 586 

\bibitem[Sch{\"o}ier et al.(2005)]{2005A&A...432..369S} Sch{\"o}ier,
  F.~L., van der Tak, F.~F.~S., van Dishoeck, E.~F., \& Black,
  J.~H.\ 2005, \aap, 432, 369

\bibitem[Sch{\"o}ier et al.(2011)]{2011A&A...530A..83S} Sch{\"o}ier,
  F.~L., Maercker, M., Justtanont, K., et al.\ 2011, \aap, 530, A83

\bibitem[Soker(2002)]{2002ApJ...568..726S} Soker, N.\ 2002, \apj, 568,
  726

\bibitem[Trams et al.(1991)]{1991A&AS...87..361T} Trams, N.~R., Waters,
  L.~B.~F.~M., Lamers, H.~J.~G.~L.~M., et al.\ 1991, \aaps, 87, 361

\bibitem[van Aarle et al.(2011)]{2011A&A...530A..90V} van Aarle, E.,
  van Winckel, H., Lloyd Evans, T., et al.\ 2011, \aap, 530, A90

\bibitem[Van Winckel et al.(1999)]{1999A&A...343..202V} Van Winckel,
  H., Waelkens, C., Fernie, J.~D., \& Waters, L.~B.~F.~M.\ 1999, \aap,
  343, 202

\bibitem[Van Winckel(2003)]{win03}
Van Winckel, H.\ 2003, ARAA, 41, 391

\bibitem[Vassilev et al.(2008)]{vassilev2008} Vassilev, V., Meledin,
  D., Lapkin, I., et al.\ 2008, \aap, 490, 1157

\bibitem[Waters et al.(1992)]{1992A&A...262L..37W} Waters, L.~B.~F.~M.,
  Trams, N.~R., \& Waelkens, C.\ 1992, \aap, 262, L37

\bibitem[Wolak et al.\ (2012)]{wolak2012} Wolak, P., Szymczak, M., \&
  G{\'e}rard, E.\ 2012, \aap, 537, A5

\bibitem[Woods et al.(2005)]{2005A&A...429..977W} Woods, P.~M., Nyman,
  L.-{\AA}., Sch{\"o}ier, F.~L., et al.\ 2005, \aap, 429, 977

\bibitem[Yang et al.(2010)]{yang10}
Yang, B., Stancil, P.C., Balakrishnan, N.; Forrey, R. C. 2010, ApJ 718, 1062


\end{thebibliography}
\end{document}